\pdfoutput=1
\documentclass{aa}
\usepackage{graphicx}
\usepackage[varg]{txfonts}

\usepackage{natbib,twoopt}

\usepackage{hyperref}
\hypersetup{breaklinks=true}
\bibpunct{(}{)}{;}{a}{}{,} %% natbib format for A&A and ApJ

\usepackage{comment}
\usepackage{pdflscape}
\usepackage[version=3]{mhchem}
\usepackage{siunitx}
\usepackage[subrefformat=parens,labelformat=parens]{subcaption}
\usepackage{enumitem}
\usepackage{indentfirst}
\usepackage{sidecap}
\usepackage{tabularx}
\usepackage{multirow}
\usepackage{lscape}
\usepackage{tablefootnote}
\usepackage{booktabs}
\usepackage{afterpage}
\usepackage{float}
\usepackage[colorinlistoftodos]{todonotes}
\usepackage{soul,xcolor}
\RequirePackage{soul}
\RequirePackage{xcolor}
\RequirePackage{todonotes}
\usepackage{stfloats}
\fnbelowfloat
\newcommand{\addColorMialy}[1]{\textcolor{black}{#1}}

\newcommand{\mialyadd}[1]{\addColorMialy{#1}}

\newcommand{\stMialy}[1]{} % Mialy's removed text not printed

\newcommand{\stZak}[1]{}
\newcommand{\Syladd}[1]{{#1}} % Sylvie's additions in black
\newcommand{\stSyl}[1]{} % Sylvie's removed "old text" not printed
\bibpunct{(}{)}{;}{a}{}{,}%% natbib format for A&A and ApJ

\newcommand{\kmps}{\ensuremath{\si{\kilo\metre\per\second}}}
\newcommand{\Kelvin}{\ensuremath{\si{\kelvin}}}
\newcommand{\figref}[1]{Fig.~{\ref{#1}}}
\newcommand{\Figsref}[1]{Figs.~{\ref{#1}}}
\newcommand{\Figureref}[1]{Figure~{\ref{#1}}}
\newcommand{\Figuresref}[1]{Figures~{\ref{#1}}}
\newcommand{\figrefb}[1]{{\ref{#1}}}

\newcommand{\MSun}{\ensuremath{M_{\odot}}}
\newcommand{\Eqref}[1]{Eq.~\eqref{#1}}
\newcommand{\Eqsref}[1]{Eqs.~\eqref{#1}}

\newcommand{\Sectref}[1]{Sect.~\ref{#1}}

\newcommand{\Sectionref}[1]{Section~\ref{#1}}
\newcommand{\Tableref}[1]{Table~\ref{#1}}

\let\oldsim\sim 
\renewcommand{\sim}{{\oldsim}}
\DeclareSIUnit\au{au}
\DeclareSIUnit\MasseSun{\MSun}
\DeclareSIUnit\yr{yr}
\DeclareSIUnit\yrs{yrs}
\DeclareSIUnit\year{yr}
\DeclareSIUnit\flowrate{\MasseSun\per\year}
\DeclareSIUnit\massdensity{\gram\per\centi\meter\cubed}

\newcommand{\orthoradius}{\ensuremath{R}}
\newcommand{\radius}{\ensuremath{r}}
\newcommand{\massflowrate}{\ensuremath{\Dot{M}}}
\newcommand{\OpeningAngleWidth}[1]{\ensuremath{\mathrm{\textbf{W}}_{\textbf{#1}}}}
\newcommand{\tracer}[1]{\ensuremath{\mathrm{\textbf{tracer}}}_{#1}}

\makeatletter
\newcommand\addcase[3]{\expandafter\def\csname\string#1@case@#2\endcsname{#3}}
\newcommand\makeswitch[2][]{%
  \newcommand#2[1]{%
    \ifcsname\string#2@case@##1\endcsname\csname\string#2@case@##1\endcsname\else#1\fi%
  }%
}
\makeatother

\makeswitch[M\_X]\modifiedrun
\addcase\modifiedrun{thetaj}{\mbox{M\_THETA}}
\addcase\modifiedrun{rhoj0}{\mbox{M\_DENSJ}}
\addcase\modifiedrun{rhoa0}{\mbox{M\_DENSA}}
\addcase\modifiedrun{DeltaV}{\mbox{M\_VARAMP}}
\addcase\modifiedrun{P}{\mbox{M\_PER}}
\addcase\modifiedrun{sawtooth}{\mbox{M\_SAWT}}
\addcase\modifiedrun{Rj}{\mbox{M\_RAD}}

\begin{document} 

   \title{Wide-angle protostellar outflows driven by narrow jets in stratified cores}
   \author{M. Rabenanahary \inst{1}
          \and
          S. Cabrit
          \inst{1,2}
          \and 
          Z. Meliani
          \inst{3}
          \and 
          G. Pineau des Forêts
          \inst{1,4}}
   \institute{Observatoire de Paris, PSL University, Sorbonne Université, CNRS, LERMA, 75014 Paris, France
   \and
   Université de Grenoble Alpes, CNRS, IPAG, 38000 Grenoble, France
   \and
   Observatoire de Paris, PSL University, Université de Paris, CNRS, LUTH  5 Place Jules Janssen, F-92190 Meudon, France
   \and
   Université Paris-Saclay, CNRS, Institut d’Astrophysique Spatiale, 91405 Orsay, France}

   \date{Received 17 January 2022 / Accepted 31 March 2022}
  \abstract{Most simulations of outflow feedback on star formation are based on the assumption that outflows are driven by a wide angle "X-wind," rather than a narrow jet. However, the arguments initially raised against pure jet-driven flows were based on steady ejection in a uniform medium, a notion that is no longer supported based on recent observations.
  We aim to determine whether a pulsed narrow jet launched in a density-stratified, self-gravitating core could reproduce typical molecular outflow properties, without the help of a wide-angle wind component.
  We performed axisymmetric hydrodynamic simulations using the MPI-AMRVAC code with optically thin radiative cooling and grid refinement down to 5~au, on timescales up to $\SI{10000}{\yrs}$. Then we computed the predicted properties for the purposes of a comparison with observational data.
  First, the jet-driven shell expands much faster and wider through a core with steeply decreasing density than through an uniform core.
  Second, when blown into the same singular flattened core, 
 a jet-driven shell shows a similar width as a wide-angle wind-driven shell
  in the first few hundred years, but a decelerating expansion on long timescales.
  The flow adopts a conical shape, with a sheared velocity field along the shell walls and a base opening angle reaching up to $\alpha \simeq 90\degr$.
  Third, at realistic ages of $\sim\SI{10000}{\yrs}$, a pulsed jet-driven shell shows fitting features along with a qualitative resemblance with recent observations of protostellar outflows with the Atacama Large Millimeter Array, such as HH46-47 and CARMA-7. In particular, similarities can be seen in the shell widths, opening angles, position-velocity diagrams, and mass-velocity distribution,
  with some showing a closer resemblance than in simulations based on a wide-angle "X-wind" model. Therefore, taking into account a realistic ambient density stratification in addition to millenia-long integration times is equally essential to reliably predict the properties of outflows driven by a pulsed jet and to confront them with the observations.}
  \titlerunning{Wide-angle protostellar outflows driven by jets in stratified cores}
   \maketitle

\section{Introduction}

The most spectacular, and often the first observed, signature 
of the birth of a new star is the formation of a slow bipolar outflow of molecular gas. This phenomenon starts in the early protostellar phase of stellar mass assembly (Class 0), persists during the envelope dispersion phase (Class 1), and is ubiquitous across all masses \citep[for a review, see e.g.,][]{Frank14}. 
Given their large sizes and high mass and momentum fluxes, ubiquity, and duration, molecular outflows are believed to play a key role in star formation on both small and large scales: recent numerical simulations \citep[see e.g.,][for a review]{Krumholtz-Federrath2019} suggest that they could be the main feedback agent setting the final stellar mass and core-to-star efficiency (via the removal of parent core material), and regulating the IMF peak, multiplicity fraction, and star formation efficiency at cluster scales (via the disruption of infall streams and replenishment of turbulence). 

The exact effect of outflow feedback, however, depends on the assumed structure for the "primary" protostellar wind sweeping up the slow outflow. Two wind models are currently in use, both involving a fast and dense jet along the flow axis \citep[as commonly observed in atomic or molecular tracers, see][]{Frank14} but strongly differing in the momentum injected at wider angles. 

The first and most frequently used wind prescription in feedback simulations is that of \citet[][hereafter MM99]{Matzner99}. It assumes a wide-angle wind radially expanding at a constant speed  ($\simeq$ 100 km/s) over all angles, with a steep density increase towards the axis responsible for the appearance of an axial "jet." This asymptotic structure was first derived for an "X-wind" magnetically launched from the disk inner edge \citep{Shu95} and only applies
to hydromagnetic winds launched radially from a narrow region (MM99). Towards the equator, the wind momentum flux is still a sizeable fraction of the isotropic wind case\footnote{the fraction is $1/\ln(2/\theta_0) \simeq 1/5 $ for
a "jet" collimation angle $\theta_0 \simeq 0.01$ rad, cf. Eq.~2 in MM99} and can directly impact equatorial infall.

The second wind model, motivated by more recent MHD simulations and observations, assumes that the fast axial jet is surrounded by a slower disk wind, ejected within a limited solid angle \citep{Federrath2014, Rohde2019}. Feedback is then dominated by the jet, with much lesser impact on equatorial regions than in the MM99 prescription. Time variability in the form of episodic outbursts was also shown to affect outflow feedback \citep{Rohde2019}. 

In principle, realistic MHD simulations of protostellar wind launching should provide the best wind prescription to adopt. However, the simulated wind structure depends on complex effects that are still the subject of intense research and debate, such as the magnetosphere-disk interaction, the distribution of magnetic flux retained in the disk long after its formation, the turbulent viscosity and resistivity, and non-ideal effects \citep[see e.g.,][]{Ireland2021,Ferreira2013-bflux,Bethune2017}.

An independent approach to determining the most realistic wind model for outflow feedback studies is to simulate the swept-up outflow properties on protostellar core scales $\leq 0.1$ pc (where the ambient density structure is dominated by self-gravity and not yet perturbed by cloud inhomogeneities) and see which wind model best reproduces the observed outflow shapes and kinematics.
Such a comparison was performed early on for two extremes in wind collimation: the wide-angle X-wind model 
\citep{Shu95,Matzner99} and a pure jet driving the outflow through large bowshocks \citep[][]{Masson-Chernin93,RagaCabrit93}. Successes and caveats were identified in each case, based on early observations  \citep[see][]{Cabrit97,LeeEtAl2001,Arce07} and these are briefly summarized and updated below.

First, in models of outflows driven by a wide-angle X-wind, the ambient medium is assumed to have a steep $1/r^2$ density decrease and a moderate degree of magnetic flattening. It is also assumed to mix instantaneously with shocked wind material. The swept-up shell then expands radially in a self-similar fashion that can reproduce several observed features of molecular outflows:  "Hubble-law" kinematics $V \propto z$, mass-velocity distribution with a power-law slope $\gamma \simeq -2$ (before opacity correction), and parabolic shapes with a wide base opening angle \citep{Shu91, LiEtShu1996, LeeEtAl2001, Shang06, Shang20}. An intrinsic caveat of this model, however, is that the uniform wind speed over all angles predicts much flatter internal shocks than observed in shock-excited H$_2$ along outflow axes. The observed curved H$_2$ bows requires a sharp drop of wind ram pressure away from the axis \citep[see discussions in][]{LeeEtAl2001, Arce07}.

 Second, simulations of jet-driven bowshocks, in contrast, reproduce  the curved morphology of internal shock fronts seen in H$_2$ very
well \citep[e.g.,][]{Suttner97,Volker99} as well as the associated "spur-like" features in CO \citep{Lee2000}. They can also reproduce the observed mass-velocity relations \citep{Downes-Cabrit2003,Moraghan08} and apparent "Hubble-laws" when the jet is variable and precessing \citep{Volker99,Rohde2019}.
The jet-driven model for outflows was strongly criticized, however, for predicting too highly elongated cavities on long timescales \citep{Ostriker2001}, too much overlapping blueshifted and redshifted emission over a wide range of inclinations \citep{LeeEtAl2001} as well as overly low velocities $\simeq 0.03$ km/s when the bowshock had expanded to typical outflow widths of 10,000 au \citep{Arce07}. As a result, it is commonly believed that jets alone cannot explain outflows with a wide opening angle, as reported for several evolved Class 1 protostars.

In order to combine the strengths of each model, a "dual wind" structure has been invoked with both a fast jet and a slower wide angle wind, where the latter would increasingly dominate at later times \citep{Yu99,Velusamy-Langer1998,Arce07,Zapata2014}. 
There are several good reasons, however, to reconsider pure jet-driven shells as the potential origin for molecular outflows. 

First, the critiques of \cite{Ostriker2001} and \cite{LeeEtAl2001} were based on models of jet-bowshocks in a uniform, or quasi-uniform, ambient medium 
(with at most a factor of 2 in density variation over the computational domain). In contrast, a steep radial $1/r^2$ density decrease is assumed in wide-angle wind-driven models to yield the apparent "Hubble-law" acceleration \citep{Shu91,LeeEtAl2001}. Such a decrease is expected on protostellar core scales $\le 0.1$ pc as a result of self-gravity. \cite{RagaCabrit93} and \cite{Cabrit97} showed that it could produce a wider opening angle for jet-driven shells, more similar to observed flows, a result confirmed for steady jets on $< 1000$ yr timescales by \citet{Moraghan08}.
Now that a steep density stratification is widely confirmed by observations of protostellar cores \citep[e.g.,][for HH46-47]{vanderMarel09} and that computational capabilities have greatly improved, it is important to explore the predicted effect on jet-driven shells over a broader parameter space and longer timescales than was feasible in the early study of \citet{Moraghan08}.

Second, the issue raised by \cite{Arce07} related to insufficient bowshock speed at large widths no longer applies with a pulsed jet. New internal bowshocks generated by the jet variability will replenish a slow jet-driven shell with faster material at observable speeds \citep{RagaCabrit93,Volker99}. In addition, the interaction between successive bowshocks will decrease their transverse speed, possibly alleviating the excessive blue and red overlap predicted at early times by \cite{LeeEtAl2001}. 
Recent Atacama Large Millimeter Array (ALMA) observations revealed multiple H$_2$ bowshocks along outflow axes, connected to nested CO cavities along their flanks \citep[e.g., HH212 and HH46-47 in ][]{Lee2015,Zhang16}. While some have been modeled with wind-driven shells \citep{Zhang19}, it is important to have similar predictions for nested jet bowshocks at realistic ages $\simeq 10^4$ yr. 

Third, recent observational studies show that
CO outflows are more elongated and collimated than initially thought: 
cloud-wide CO maps, as well as optical and infrared imaging surveys, show that at least 40\% of outflows are more than a parsec long \citep[][and references therein]{Frank14}.
A striking example is the B5-IRS1 outflow, driven by a Class 1 protostar. While it exhibits a (projected) full opening angle $\alpha \simeq \SI{100}{\degree}$ at its base, 
argued as evidence for a wide-angle wind,
unbiased CO maps of its parent cloud reveal that each lobe extends (at least) up to 2.2 pc from the source \citep{Arce2010}. 
The length-to-width ratio is then $q > 11$ \citep[see maps in][]{Frank14}. This is 
inconsistent with current models of shells driven by a wide angle X-wind \citep[which predict an aspect ratio $q < 3.5$ for a base opening angle $\alpha \ge 70\degr$, see Table 2 in][]{Shang20}. The parsec size of many outflows appears more suggestive of jet-driven flows.

Estimated outflow opening angles are also affected by several biases: angular resolution, the height at which they are measured \citep{Velusamy-Langer1998,Velusamy2014}, and inclination (angles appear wider in flows seen closer to pole-on). Two recent studies in Orion minimize these biases by providing uniform measurements at the same (high) linear resolution and projected height over randomly selected samples. 
In an ALMA survey of the widths of 22 (mostly Class 0) CO outflows \citep{Dutta2020}, 50\% subtend projected full-opening angles in the range $\alpha =$ 25\degr-65\degr\ at a projected altitude $z_{\rm proj}=800$~au. 
In a sample of 29 older outflow cavities (mostly Class 1) imaged in scattered light with HST \citep{Habel2021}, we see that 50\% are in the range $\alpha =$ 8\degr-46\degr\ at $z_{\rm proj}=8000$~au, and the fraction of point sources (viewed down the cavity interior) suggests a maximum deprojected opening angle $\alpha_{\rm deproj} \le 70\degr$. Therefore, CO outflows seem to be more collimated on average than previously believed and it is necessary to investigate whether a pure jet (in a stratified core) could reproduce typical observed widths, before drawing any conclusions on a dominant contribution from a wider angle wind.

Here, we examine this issue by presenting the first high-resolution simulations of pure jet-driven shells in strongly stratified cores, up to ages of $\SI{10000}{\yrs}$ and physical scales of 0.1 pc. For the first time in jet simulations, we consider the same flattened core structure as in the competing model of outflow driven by a wide-angle X-wind \citep{LiEtShu1996,LeeEtAl2001}. We show that pulsed conical jets propagating in this environment sweep up a wider outflow cavity than in a uniform medium, with a width and opening angle that are compatible with recent outflow surveys. The predicted position-velocity diagrams and mass-velocity relation also show a promising qualitative agreement with recent ALMA outflow observations at high resolution and sensitivity, without any of the caveats noted previously for jet bowshocks in uniform media. 

In Section 2, we present our numerical method and generic set up. In Section 3, we present the effect of a 1D and 2D density stratification on a jet-driven shell. In Section 4, we introduce a small jet opening angle and explore the effect of various free parameters on the cavity shape and kinematics. In Section 5, we present a simulation up to $\SI{10000}{\yrs}$ and compute predicted flow widths, position-velocity diagrams, and mass-velocity relation, finding excellent qualitative agreement with recent ALMA observations. Section 6 summarizes our main results and conclusions.

\section{General numerical setup}
\label{section:code_setup_model}

%%%%%%%%%%%%%%%%%%%%%
% Numerical methods in use
%%%%%%%%%%%%%%%%%%%%%
\subsection{Governing equations, code, and numerical method}
%%%%%%%%%%%%%%%%%%%%%
% MPI-AMRVAC Basic descriptions
%%%%%%%%%%%%%%%%%%%%%
We performed axisymmetric 2D hydrodynamic simulations  
in cylindrical coordinates $\left(\orthoradius,z\right)$, using the Message Passing Interface-Adaptive Mesh Refinement Versatile Advection Code \citep[MPI-AMRVAC; ][]{Keppens21}. The hydrodynamics module of this finite volume, cell-centered code solves the hydrodynamic equations of mass, momentum, and energy conservation described, respectively, by :
%%%%%%%%%%%%%%%%%%%%%
% HD EQUATIONS
%%%%%%%%%%%%%%%%%%%%%
 \begin{eqnarray}\label{Eq:S_EQ_HD_conserved}
 \frac{\partial \rho}{\partial t}~+~\nabla\cdot(\rho\mathbf{v})~ &=& 0 \,, \label{eq:euler:mass}\\
            \frac{\partial (\rho\mathbf{v})}{\partial t}~+~\nabla\cdot(\rho\mathbf{vv})~+\nabla p ~ &=&  \mathbf{F}_p\,,\label{eq:euler:momentum} \\
            \frac{\partial e}{\partial t}~+~\nabla\cdot(e\mathbf{v}+ p\mathbf{v})~  &=& - %\left(\frac{\rho}{\mu m_{H}}\right)^2 \Lambda(T)\,, \label{eq:euler:energy}    
            n_H^2 \Lambda(T)\,, \label{eq:euler:energy}    
 \end{eqnarray}
%%%%%%%%%%%%%%%%%%%%%
% Physical variables description
%%%%%%%%%%%%%%%%%%%%%           
where $\rho$ is the mass density, $\mathbf{v}$ is the velocity vector, $p$ is the thermal pressure, and $e=p/(\gamma-1)+\rho\,\mathbf{v}^2/2$ is the total (thermal and kinetic) energy density, with 
$\gamma$ the adiabatic index (taken here as $5/3$). Two source terms are introduced on the right-hand side: 
following \cite{LeeEtAl2001}, an inward-directed force field, $\mathbf{F}_p = \mathbf{\nabla} p(t=0)$, is imposed to maintain the unperturbed stratified ambient core in hydrostatic equilibrium at any time. In addition, optically thin equilibrium radiative cooling is included as a source term $-n_H^2 \Lambda(T)$ in the energy equation, 
with $n_H$ the number density of hydrogen nuclei and $T$ the gas kinetic temperature \citep{VanMarle11}.
%%%%%%%%%%%%%%%
%nH and He fraction
%%%%%%%%%%%%%%%
We consider an atomic gas with 
a standard helium fraction $x({\rm He}) = n({\rm He})/n_H=0.1$, such that $n_H=\rho/(1.4 m_H)$ with $m_H$ the mass of a proton.
The temperature is inferred from gas pressure using the perfect gas law, $T = p/(n_{\rm tot} k_B),$ with
$n_{\rm tot}$ as the total number of particles per unit volume.
Given the moderate shock speeds encountered in our simulation, we assume that hydrogen and helium remain mostly neutral, so that $n_{\rm tot} \simeq n({\rm H}) + n({\rm He}) = 1.1 n_H$. 
%%%%%%%%%%%%%%%%%%%%%
% Physic in use concerning cooling
%%%%%%%%%%%%%%%%%%%%%
The cooling curve $\Lambda(T)$ depends on the local temperature, $T$, and the gas metallicity.
%%%%%%%%%%%% Cooling table in use
Here, we use an atomic cooling curve with solar metallicity.
Aside from \Sectref{section:PCJ} (where we use the same cooling function as \citet{LeeEtAl2001} for comparison purposes), we use the cooling function $\Lambda(T)$ from \citet{Schure09} throughout the study, 
which  combines 
the equilibrium cooling curve 
generated with the SPEX code \citep{KaastraandNewe2000} for temperatures above $10^4$ K,
and the cooling curve from \cite{DalgarnoEtMcCray72} with an ionization fraction $x_{ion}=10^{-3}$ 
for temperatures below $10^4$ K.
%%%%%%%%%%%% TEMPERATURE FLOOR
In each simulation, the minimum temperature for radiative cooling is set at the initial ambient core temperature.
This prevents the non-realistic radiative cooling of unshocked material inside the jet beam and the ambient core.
        
%%%%%%%%%%%%%%%%%%%%%
% Numerical methods in use
%%%%%%%%%%%%%%%%%%%%%
 To solve \Eqsref{eq:euler:mass}-\eqref{eq:euler:energy}, we chose a Harten–Lax–van Leer contact (HLLC) scheme \citep{Li05}  with a minmod limiter. This combination is extremely robust in handling the shocks encountered in our problem.
 
As for the boundary conditions, the jet axis ($R=0$) is treated as an axisymmetric boundary; the boundary conditions at $z=0$  are fixed inside the jet inlet ($R < R_j$) and open in the surrounding ambient core.
%%% OUTER BOUNDARIES
The outer limits of the computational domain are treated as open boundaries.
%%%%%%%%%%%%%%%%%%%%%
% Inlet definition
%%%%%%%%%%%%%%%%%%%%%
The jet inlet (inside which the density and velocity vectors are prescribed at each time step) is a conical domain of radius $R=\orthoradius_j$ at $z=0$, height $z_i$, and semi-open angle $\theta_j$.

%%%%%%%%%%%%%%%%%%%%%%%
% AMR STRATEGY
%%%%%%%%%%%%%%%%%%%%%%%
Due to the high dynamic ratio between inner jet shock scale and propagation scales studied in this paper, it is crucial to use adaptive mesh refinement (AMR) to resolve the shocks in a cost-effective manner. 
Apart from \Sectref{section:PCJ} (where we use the same fixed grid resolution as \citet{LeeEtAl2001} for comparison purposes), all simulations performed in this paper have a base grid that is allowed to be locally refined up to four times, doubling the resolution at each new level of refinement (that is, with a highest grid resolution at level 5 that is $2^4$ finer than at level $1$).  The refinement-derefinement criterion is based on a Lohner error estimator on the quantity
$n_H^2\Lambda(T)$.
We further ensure that the highest level of refinement is always present inside the jet beam.

\subsection{Velocity variability of the jet}

Pulsed jets are simulated by time-varying \mialyadd{the jet velocity} around its initial value $v_0$, as

        \begin{equation}
            v_j(t)=\left|\left|\mathbf{v_j}(t)\right|\right| = v_0 
            h(t)\,, \label{eq:vjt}
        \end{equation}
        
\noindent where $h\left(t\right)$ is the dimensionless variability profile defining how the jet will pulse. 

Following \citet{LeeEtAl2001}, we keep the jet mass-flux constant over time in the present paper, by imposing an inverse variation of the jet density: 
       \begin{equation}
            \rho_{j0}(t) = \frac{\rho_{j0}}{h\left(t\right)} = \frac{\rho_{j0}v_0}{v_j(t)},  
            \label{eq:rhojt}
        \end{equation}
        \noindent where $\rho_{j0}$ is \mialyadd{the initial jet density at $z=0$, $R = R_j$, and $t=0$.}

\subsection{Tracers}

%%%%%%%%%%%%%%%%%%%%%
% Tracers
%%%%%%%%%%%%%%%%%%%%%
Following \cite{Porth14}, the code also solves the advection equation of two passive fluid tracers, $\tracer{a}$ and $\tracer{j}$, which trace fluid parcels originating from the ambient core and from the jet, respectively. They evolve according to:
\begin{align}
    \frac{\partial \tracer{k}}{\partial t}~+~\mathbf{v}\cdot\nabla~\tracer{k} &= 0\,\mbox{, for }k=a,j\,. \label{eq:tracers_dyn}\
 \end{align}
These tracers are dimensionless numerical quantities.
To improve numerical precision and accurately track mixing between the jet and ambient core material, they are taken to cover a wide range $\left[0,\,10^{7}\right]$.
 Thus, at $t=0$, $\tracer{j}$ (resp. $\tracer{a}$) is initialized to $10^7$ inside the jet beam (resp. the surrounding core)
 and is set to zero elsewhere.
 
 From those tracers, we may compute the local fraction of material originating from the ambient core and from the jet, $f_a$ and $f_j$, inside each mesh cell and at any time step, as

 \begin{align}
     f_{k}= \frac{\tracer{k}}{\tracer{a}+\tracer{j}},\,\mbox{for }k=a,j\,.
            \label{eq:tracers_fractions}
        \end{align}

\section{Pulsed cylindrical jet in a non-uniform medium}
\label{section:PCJ}

In this section, we investigate how the shape of the shell driven by the same cylindrical jet as in \cite{LeeEtAl2001} is influenced by a more realistic, steep density decrease in the ambient core  (\Sectref{section:PCJ:results}). We then compare with the wide-angle wind-driven shell modeled in \cite{LeeEtAl2001} for the same age and core stratification (\Sectref{section:PCJ:pcj_vs_waw}).

\subsection{Setup and choice of density  stratifications}
\label{section:PCJ:setup}

\begin{table*}
\caption{Parameters of the pulsed cylindrical jet simulations (\Sectref{section:PCJ})}
\label{table:PCJ:parameters}
\begin{center}
\begin{minipage}{\textwidth}
\begin{tabular}{l |l l l l}
 \hline
  \hline
     \multicolumn{5}{c}{\textit{Fixed parameters (from Lee et al. 2001)}}
    \\
     \hline
     one-sided jet mass-loss rate  &
     \multicolumn{4}{l}{$\massflowrate = \SI{6.0e-8}{\MSun/yr}$}
     \\
     jet semi-opening angle  & 
     \multicolumn{4}{l}{$\theta_j = 0\degr$}
     \\
     jet radius  & 
     \multicolumn{4}{l}{$R_j = \SI{2.5e15}{\cm}$}
     \\
     mean jet velocity  &
     \multicolumn{4}{l}{$v_0 = \SI{120}{\kmps}$}
     \\
     jet velocity variation &
     \multicolumn{4}{l}{$v_j(t) = v_0 + {\Delta V} \sin {\frac{2\pi t}{P}}$}
     \\ 
     semi-amplitude  &
     \multicolumn{4}{l}{$\Delta V = \SI{60}{\kmps}$}
     \\
     jet density variation  &
     \multicolumn{4}{l}{$\rho_j(t) = \rho_{j0}\,[v_0 / v_j(t)]$ \quad (constant mass-flux)}
     \\
     jet density at $t=0$ & 
     \multicolumn{4}{l}{$\rho_{j0} = \SI{1.6e-20}{\gram\per\centi\meter\cubed}$}
     \\
     jet temperature  &
     \multicolumn{4}{l}{$T_j$ = $\SI{270}{\kelvin}$}
     \\
     ambient core temperature  &
     \multicolumn{4}{l}{$T_a = \SI{30}{\kelvin}$}
     \\
     radiative cooling function  &
     \multicolumn{4}{l}{$\Lambda\left(\SI{270}{\Kelvin}\leq T <10^4~\SI{}{\Kelvin}\right)$ from \cite{DalgarnoEtMcCray72}}
     \\
     & 
     \multicolumn{4}{l}{$\Lambda\left(T\geq10^4~\SI{}{\Kelvin}\right)$ from \cite{MacDonald81}}
     \\
     simulation domain  &
     \multicolumn{4}{l}{$\left(\orthoradius,z\right)
     =
     \left(\SI{7.0e16}{\cm},\SI{1.4e17}{\cm}\right)
     =
     \left(\SI{4679}{au},\SI{9358}{au}\right)$ }
     \\
     number of cells  &
     \multicolumn{4}{l}{$n_\orthoradius\times n_z
     =
     336\times672$ }
     \\
     resolution  &
     \multicolumn{4}{l}{$\Delta\orthoradius=\Delta z
     =
     \SI{2.08e14}{\cm}=
     \SI{14}{au}$}
     \\
     \hline
     \hline
     \multicolumn{5}{c}{\textit{Variable parameters}}
     \\
     \hline
     model name & PCJ-U$^{\mathrm{a}}$ & PCJ-Z & PCJ-R & 
     \multicolumn{1}{|l}{PCJ-RW$^{\mathrm{b}}$} 
    \\
     & 
     (\figref{fig:PCJ}a) &
     (\figref{fig:PCJ}b) &
     (\figref{fig:PCJ}c) &
    \multicolumn{1}{|l}{(\figref{fig:PCJ-RW})} 
     \\     
     \hline
     core density profile & uniform & $z$-stratified & \multicolumn{2}{|l}{flattened singular core} \\
     $\rho_a(\mathbf{r})$ &
     $\rho_{a0}$ &
     ${\rho_{a0}}\left(1+\frac{z}{z_c}\right)^{-2}$ &
     \multicolumn{2}{|l}{$\rho_{a0}\sin^2{\theta}$ ${\radius_0^2}\left({\orthoradius^2+z^2}\right)^{-1}$}
     \\
      &
      &
     $z_c = 10^{16}$ cm &
     \multicolumn{2}{|l}{$\radius_0 = 2.5 \times 10^{15}$ cm}
     \\
     \hline
     core density at $z=0,R=\SI{2.5e15}{\cm}$ &
     \multicolumn{3}{l|}{$\rho_{a0} = \SI{1.6e-20}{\gram\per\centi\meter\cubed}$} & {$\SI{1.6e-18}{\gram\per\centi\meter\cubed}$}
     \\
     core/jet density \stSyl{ratio} \Syladd{contrast at $z=0$}  & 
     \multicolumn{3}{l|}{$\eta={\rho_{j0}}/{\rho_{a0}}= 1$} & 0.01
     \\
     jet variability period  &
     \multicolumn{3}{l|}{$P = \SI{310}{\yrs}$} & $\SI{115}{\yrs}$
     \\ 
     simulation age &
     \multicolumn{3}{l|}{$t = \SI{610}{\yrs}$} & $\SI{296}{\yrs}$
     \\ 
     \hline
\end{tabular}
\begin{list}{}{}
\item[$^{\mathrm{a}}$] identical parameters to pulsed jet simulation of Lee et al. (2001) (see their Fig. 7)
\item[$^{\mathrm{b}}$] identical parameters to pulsed wind simulation of Lee et al. 2001 (cf. their Fig. 12), except that the wide-angle wind is replaced here by a cylindrical jet of same mass-flux, injection radius, and velocity variability.
\end{list}
\end{minipage}
\end{center}
\end{table*}

To make a  comparison with \cite{LeeEtAl2001}, we 
adopted their pulsed cylindrical jet model (thereafter PCJ) with the same
jet radius, $\orthoradius_j$, temperature, $T_j$, sinusoidal variation of jet velocity, $v_j(t)$, inversely varying 
jet density, $\rho_j(t)$, and constant one-sided jet mass-loss rate, $\massflowrate$. 
We also adopted the same uniform spatial resolution of 14 au, 
radiative cooling law (using \citet{MacDonald81} above $10^4$~K), domain size, ambient temperature, and simulation ages. 
The particular setup presented in this section is summarized in \Tableref{table:PCJ:parameters}. 

We then investigated three different ambient core stratifications: first, just as in the jet-driven simulations of \cite{LeeEtAl2001}, we considered a uniform core, with  
        \begin{equation}
            \rho_a(\orthoradius,z) = ~\rho_{a0}\,. 
            \label{profiles:density:uniform}
        \end{equation}
Second, we investigated a $z$-stratified core following the prescription in \cite{Cabrit97}:

        \begin{align}
        \rho_a(\orthoradius,z) = ~{\rho_{a0}} \left(1+\frac{z}{z_c}\right)^{-2}
        \quad {\rm with} \quad  z_c = 10^{16} ~{\rm cm}\,,
        \label{profiles:density:cabrit97}        \end{align}    
\noindent where the value of $z_c$ is justified by observations of protostellar cores showing a steep stratification from $\SI{20000} {au}$ down to (at least) $\SI{1000}{au}$ \citep[see e.g.,][]{Motte2001,vanderMarel09}. \cite{LeeEtAl2001} investigated (see their Section 3.6) a somewhat similar $z$-stratification $\rho_a(z) \propto 1/[ 1+({z}/{z_c})^2]$ but with a  flattening scale $z_c = \SI{1.25e17}{\cm}$ that is ten times greater than in \Eqref{profiles:density:cabrit97} and equal to the size of their computational box. The ambient density thus dropped only by a factor of 2 across the whole domain and 
no strong effect on shell morphology was reported.

\begin{figure*}[!tp]
    \centering
    \includegraphics[trim=3.5cm 3.0cm 0 0, clip,width=\textwidth]{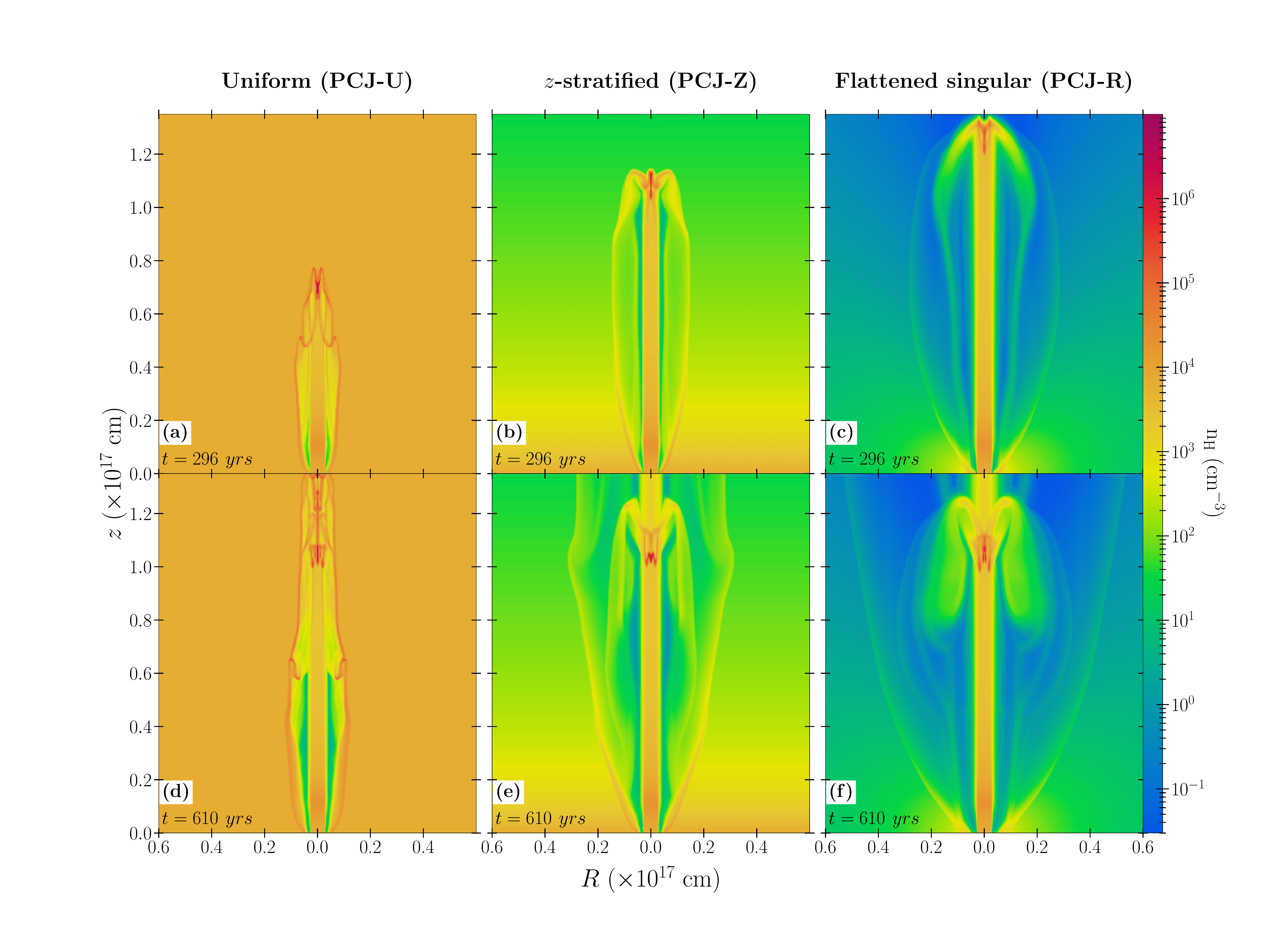}
    \caption{Maps of hydrogen nuclei density $n_H$ from simulations of the same cylindrical jet as in \cite{LeeEtAl2001} (see \Tableref{table:PCJ:parameters}). The three columns confront three different ambient core stratifications, and the two rows show the map at two ages $t=\SI{296}{\yrs}$ \textit{(top)} and $\SI{610}{\yrs}$ \textit{(bottom)}. The jet is launched in \figref{fig:PCJ}a and \figrefb{fig:PCJ}d through a uniform ambient core with profile $\rho(\orthoradius,z)=\rho_{a0}$; in \figref{fig:PCJ}b and \figrefb{fig:PCJ}e through an $z$-stratified ambient core $\rho(\orthoradius,z)=\rho_{a0}/(1+z/z_c)^2$ with $z_c =10^{16}\SI{}{\cm}$; and in \figref{fig:PCJ}c and \figrefb{fig:PCJ}f through a flattened singular core $\rho(\radius,\theta)=\rho_{a0}\sin^2{\theta}(\radius_0/\radius)^2$, where $r$ is the spherical radius and $\radius_0=\SI{2.5e15}{\cm}$. 
    All core density profiles have the same base density at $R = R_j$, $\rho_{a0} =\SI{1.6e-20}{\gram\per\centi\meter\cubed}$, density-matched with the jet at $t=0$. W note how the jet-driven shell expands faster and wider through an increasingly stratified core, whereas the nested shells also grow wider.} 
    \label{fig:PCJ}
\end{figure*}

Third, we considered the same flattened singular core profile 
as in the wide angle wind-driven simulations of \cite{LeeEtAl2001}, namely:
\begin{align}
             \rho_a(\radius,\theta) = ~ \rho_{a0}\sin^2{\theta}\left(\frac{\radius_0^2}{\radius^2}\right) 
              \quad {\rm with} \quad  \mialyadd{\radius_0} =  2.5 \times 10^{15} {\rm cm},
            \label{profiles:density:lee2001}
\end{align}   
where $\radius = (\orthoradius^2+z^2)^{1/2}$ denotes the spherical radius and
$\theta$ is the polar angle from the jet axis. 
This $1/r^2$ decrease is appropriate for a self-similar singular isothermal core. The $\sin^2 \theta$ 
dependence is an approximation for the flattened
magnetostatic equilibrium solution of \cite{LiEtShu1996} with $n = 2$ (where this parameter measures the degree of magnetic support and equatorial flattening). 
When a core is stratified according to this particular solution and is then swept-up by an X-wind, it can reproduce the typical collimation, parabolic shape, and mass-velocity distribution in CO outflows, while also being consistent with the mean observed flattening of prestellar cores
\cite{LiEtShu1996}.  Hence, analytical models and numerical simulations 
of X-wind driven shells \citep{Lee2000,LeeEtAl2001,Zhang19} widely adopt the expression in \Eqref{profiles:density:lee2001} as a "standard" ambient medium.
However, to our knowledge, it was never used in jet-driven outflow simulations until now.

\subsection{Effect of ambient core stratification on jet-driven shells}
\label{section:PCJ:results}

\Figureref{fig:PCJ} shows the results of our simulations at the same ages and for 
the same cylindrical jet propagating into three different density distributions: uniform 
(model PCJ-U, \figref{fig:PCJ}a and d),
$z$-stratified (model PCJ-Z, \figref{fig:PCJ}b and e),
and flattened singular core (model PCJ-R, \figref{fig:PCJ}c and f). 
All three have the same value of $\rho_{a0} = \SI{1.6e-20}{\gram\per\centi\meter\cubed}$ in \Eqsref{profiles:density:uniform}, \eqref{profiles:density:cabrit97}, and \eqref{profiles:density:lee2001}, respectively. Since $R_j = \SI{2.5e15}{\cm} = r_0$, the ambient density at the jet base $(z=0,R=R_j)$ is also the same in all three cases (equal to $\rho_{a0}$), as well as the initial ($t=0$) jet-ambient density contrast $\eta = 1$ at this point. 
All  model parameters are summarized in \Tableref{table:PCJ:parameters}.

Going from left to right in \figref{fig:PCJ}, the ram pressure constraints exerted on both the main and nested shells are relaxed along the \mialyadd{$z$-axis, and then along both the radial and polar directions.} As we may see, this change leads to a main shell expanding faster along $z$ and wider over time. 

As already noted by \cite{LeeEtAl2001}, we confirm that a uniform medium produces a narrow and roughly cylindrical jet-driven shell, unlike observed outflows.
However, introducing a steep $1/z^2$ stratification here leads to a more conical leading shell as time proceeds, confirming the analytical and numerical predictions of \cite{RagaCabrit93} and \cite{Cabrit97}. 
For example, at $\SI{610}{\yrs}$, the shell full width at 
the top of the domain
is increased by a factor 4 between PCJ-U and PCJ-Z (from $\SI{1.3e16}{\cm}$ to $\SI{5.6e16}{\cm}$).
In the flattened singular core PCJ-R, the rarefied polar holes create an even wider shell with a parabolic shape and a twice larger full width than in PCJ-Z ($\SI{1.0e17}{\cm}$).

Even though the ambient density at the jet inlet $(z = 0, \orthoradius = \orthoradius_j)$ was kept the same, a steeper density stratification also increases the shell full width near its base. From \figref{fig:PCJ}, we can measure full widths at $z$ = 800 au (= $\SI{1.2e16}{\cm}$) of
$\OpeningAngleWidth{800} = \SI{0.95e16}{\cm}$, $\SI{1.16e16}{\cm}$ and $\SI{1.75e16}{\cm}$
for the uniform medium, $1/z^2$ decrease and flattened singular core, respectively.

In addition to the leading shell carved by the supersonic jet head, 
jet variability produces successive and periodic internal working surfaces (thereafter IWS) where high-pressure shocked material is ejected sideways, forming bowshocks expanding inside the leading shell and producing nested "secondary" shells, visible in the $t=\SI{610}{\yrs}$ snapshot in \figref{fig:PCJ}.
As the leading shell expands faster and wider in a stratified core, its inner density distribution (and, hence, the pressure locally exerted on the nested shells) drops more steeply than through an uniform core. This allows the nested shells to expand more widely as well. In parallel, each IWS still moves along the jet beam at the same velocity, independently of the core stratification. This is because the IWS propagation speed only depends on the jet velocity and density conditions upstream and downstream of the forming working surface \citep{Raga90}.
These conditions are entirely determined by the jet variability properties, which remain unmodified for each of our core density profiles.

\subsection{Comparison with a wide-angle wind-driven shell}
\label{section:PCJ:pcj_vs_waw}

Here, we adopt the same setup as in PCJ-R (\figref{fig:PCJ}c,f), but with a flattened singular core that is 100 times denser
and identical to that considered in the wide-angle wind models of \cite{LeeEtAl2001}, with $\rho_{a,0} = \SI{1.6e-18}{\gram\per\centi\meter\cubed}$. We also adopt a shorter variability period $P=\SI{115}{\yrs}$ than in \Sectref{section:PCJ:results}, so that our pulsed cylindrical
jet has the exact same mass-loss rate, injection radius, and velocity variability as their pulsed wide-angle wind.
The parameters of this new model (PCJ-RW) are summarized in the last column of \Tableref{table:PCJ:parameters}. 

\begin{figure}[h]
    \resizebox{\hsize}{!}{\includegraphics[trim=0.5cm 1.5cm 0 0,clip]{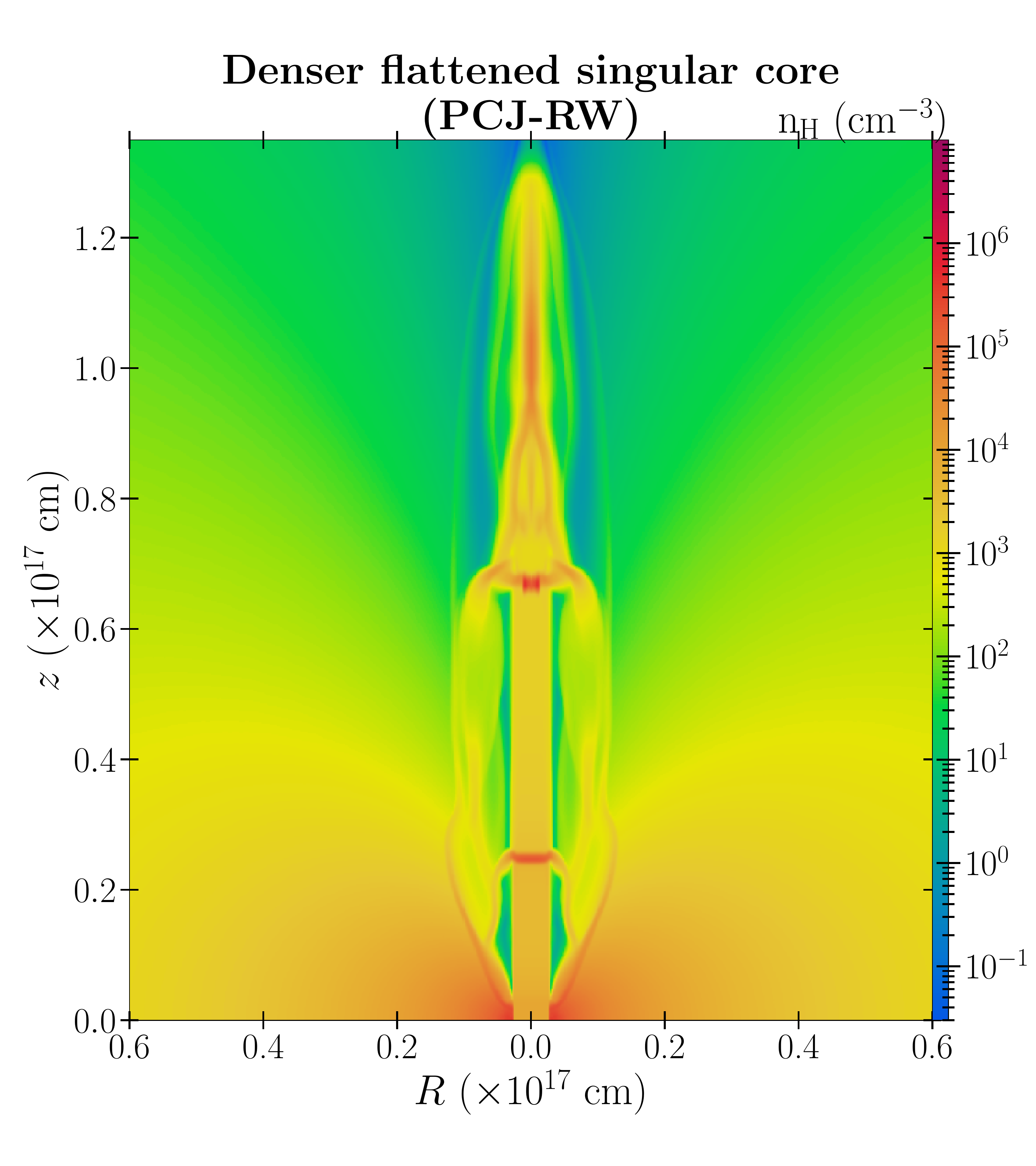}}
    \caption{ Density snapshot at $t=\SI{296}{\yrs}$ of pulsed cylindrical jet model PCJ-RW, with same injected mass-loss rate, velocity variability, and ambient density distribution as the pulsed wide-angle wind model in Fig. 12 of \cite{LeeEtAl2001}. The resulting shell size and opening angle are the same as for the wind-driven shell, at this young age.}
    \label{fig:PCJ-RW}
\end{figure}

\Figureref{fig:PCJ-RW} shows the resulting density map of the PCJ-RW simulation 
    at an age $t=\SI{296}{\yrs}$. This map can directly be compared with the wide-angle wind simulation at the same age in Fig.~12 of \cite{LeeEtAl2001}. 
     At this early age, the jet-driven shell opens as wide as for the pulsed wide-angle wind in the same ambient stratification; namely, we measure similar maximum shell widths of 
     $\SI{2.6e16}{\cm}$ ($\sim 10R_j$) for both models, and a similar full opening angle\footnote{Defined following \cite{Dutta2020} as $\alpha_{800} = 2\arctan(\OpeningAngleWidth{800}/[2\times \SI{800}{au}])$.} at $z =\SI{800}{au}$ of $\alpha_{800}=\SI{70}{\degree}$  for PCJ-RW and  $\alpha_{800}=\SI{80}{\degree}$ for the wide-angle wind model.

\subsection{Summary}

The usual criticism of jet-driven shells producing overly narrow opening angles appears no longer valid when a realistic stratified ambient medium is considered.
The morphology of the shell driven by a pulsed jet is strongly affected by a steep stratification in density of the ambient core surrounding the jet. Wider shells are formed with increased opening angle near the base of the outflow and conical or parabolic shapes on large scales are highly reminiscent of observed CO outflows. Furthermore, in a standard flattened singular core, the jet-driven shell initially expands as wide as with a wide-angle wind.
In the rest of this paper, we examine which factors affect the jet-driven shell shape (see \Sectref{section:monovariated_run}) and how it evolves on longer timescales $\sim\SI{10000}{yrs,}$ which more closely resemble the actual outflow ages (see \Sectref{section:long_term}).

\section{Shells driven by a conical jet in a flattened singular core}
\label{section:monovariated_run}

\begin{table*}
\caption{Parameters of pulsed conical high-density jet simulations with the resulting opening angles and full widths (\figref{fig:monovariated_run:maps})}
\label{table:monovariated_run:parameters}
\begin{center}
\begin{minipage}{\textwidth}
\begin{tabular}{l l l l l l l}
 \toprule
  \hline
     \multicolumn{7}{c}{\textit{Fixed parameters}}
     \\
     \hline
     mean jet velocity  &
     \multicolumn{6}{l}{$v_0 = \SI{120}{\kmps}$}
     \\     jet density variation  &
     \multicolumn{6}{l}{$\rho_j(t) = \rho_{j0}\,[v_0/ v_j(t)]\times(\orthoradius_j^2+z_0^2)(\orthoradius^2+[z+z_0]^2)^{-1}$,  with $z_0=\orthoradius_j/\tan\theta_j$\quad (constant mass-flux)}
     \\
     core density profile & \multicolumn{6}{l}{Flattened singular core
     $\rho_a(\mathbf{r})=\rho_{a0}\sin^2{\theta} \,{\radius_0^2}\,\left({\radius^2}\right)^{-1}$, with $\radius_0=\SI{2.5e15}{\cm}$}
     \\
     jet temperature  &
     \multicolumn{6}{l}{$T_j$ = $\SI{100}{\kelvin}$}
     \\
     ambient core temperature  &
     \multicolumn{6}{l}{$T_a = \SI{100}{\kelvin}$}
     \\
     radiative cooling function  &
     \multicolumn{6}{l}{$\Lambda\left(\SI{100}{\Kelvin}\leq T <10^4~\SI{}{\Kelvin}\right)$ from \cite{DalgarnoEtMcCray72}}
     \\
     & 
     \multicolumn{6}{l}{$\Lambda\left(T\geq10^4~\SI{}{\Kelvin}\right)$ from \cite{Schure09}}
     \\
     simulation domain  &
     \multicolumn{6}{l}{$\left(\orthoradius,z\right)
     =
     \left(\SI{7.0e16}{\cm},\SI{1.9e17}{\cm}\right)=\left(\SI{4679}{au},\SI{12700}{au}\right)$}
     \\
     number of cells  &
     \multicolumn{6}{l}{$n_\orthoradius\times n_z
     =
     56\times152$ for the full grid at AMR level 1}
     \\
    maximum resolution  &
     \multicolumn{6}{l}{$\SI{7.8e+13}{\cm}=\SI{5.2}{au}$ at AMR level 5}
     \\
     snapshot age  & \multicolumn{6}{l}{$\SI{700}{\yrs}$} 
     \\
     \toprule\hline
       \multirow{2}{1.0cm}{Parameter} & 
       \multirow{2}{2.5cm}{Reference Model H\_REF} & 
       \multirow{2}{2.1cm}{Modified parameter 
       \footnote{In each modified model, only one parameter at a time is changed with respect to the reference model.}}&
       \multirow{2}{2.1cm}{Modified Model name}
      & $\alpha_{800}$\footnote{Shell full opening angle at $z = 800$ au. We obtain $\alpha_{800}=\SI{88}{\degree}$ for the reference model.} &
      
      $\OpeningAngleWidth{\SI{800}{}}$\footnote{Full shell width at $z = 800$ au. We obtain $\OpeningAngleWidth{\SI{800}{}}=\SI{2.3e16}{\cm}$ for the reference model.} &
      $\OpeningAngleWidth{\SI{12700}{}}$\footnote{Full shell width at $z = 12700$ au (top of the grid). We obtain $\OpeningAngleWidth{\SI{12700}{}}=\SI{7.6e16}{\cm}$ for the reference model.}
     \\ 
        & 
        & 
       &  
      & ($\SI{}{\degree}$)
      & ($10^{16}~\SI{}{\cm}$)
      & ($10^{16}~\SI{}{\cm}$)
     \\ 
     \hline
     core base density $\rho_{a0}$ &
     $\SI{1.6e-18}{\gram\per\centi\meter\cubed}$ &      $\SI{1.6e-20}{\gram\per\centi\meter\cubed}$ 
     &
     H\_DENSA &  $\SI{112}{}$
     & $\SI{3.5}{}$
     & $\SI{13.8}{}$
     \\
     jet semi opening angle  $\theta_j$ &
     $\SI{3}{\degree}$ & $\SI{7}{\degree}$ &
     H\_THETA 
     & $\SI{95}{}$
     & $\SI{2.6}{}$
     & $\SI{10.5}{}$ \\
       jet base initial density\footnote{High jet density case, yielding a one-sided mass-flux 
       $\massflowrate = \SI{6e-6}{\MSun/yr}$ for the reference values of $\theta_j$ and $R_j$}  $\rho_{j0}$&
      $\SI{1.8e-17}{\gram\per\centi\meter\cubed}$ & $\SI{1.8e-15}{\gram\per\centi\meter\cubed}$ &
     H\_DENSJ & $\SI{86}{}$ 
     & $\SI{2.2}{}$
     & $\SI{8.4}{}$
     \\
          semi-amplitude $\Delta V$ &
     $\SI{60}{\kmps}$ & $\SI{90}{\kmps}$ &
     H\_VARAMP & $\SI{91}{}$ 
     & $\SI{2.4}{}$
     & $\SI{8.0}{}$
     \\          
     jet variability period $P$  &
     $\SI{115}{\yrs}$ & 
     $\SI{300}{\yrs}$ &
     H\_PER &  $\SI{90}{}$ 
     & $\SI{2.4}{}$
     & $\SI{7.4}{}$
     \\ 
     jet variability profile $h(t)$&
      $1+ \frac{\Delta V}{v_0} \sin {\frac{2\pi t}{P}}$ & 
      $1+\frac{\Delta V}{v_0}
      \left\{1 - \right.$
      & \multirow{2}{0.5cm}{H\_SAWT} & \multirow{2}{0.5cm}{$\SI{86}{}$} 
      &\multirow{2}{0.5cm}{$\SI{2.2}{}$}
      &\multirow{2}{0.5cm}{$\SI{7.2}{}$}
     \\
     &
       & 
      $\left.2\cdot \mathrm{mod}\left(\frac{t}{P},1\right)\right\}$
      &  & &
     \\
     jet radius $\orthoradius_j$ &
      $\SI{7.5e14}{\cm}$ & $\SI{3.0e14}{\cm}$
       & 
     H\_RAD & $\SI{84}{}$ 
     & $\SI{2.2}{}$
     & $\SI{6.9}{}$
     \\
\toprule
\end{tabular}
\end{minipage}
\end{center}
\end{table*}

In this section, we investigate the influence that several physical parameters (expected to vary among observed sources) have on the morphology and kinematics of jet-driven shells.
For consistency, we keep the singular flattened core stratification in Equ. \ref{profiles:density:lee2001}
(as in wide-angle wind-driven models),
but we adopt a more realistic conical jet geometry. 
High-resolution jet observations suggest jet half-opening angles $\theta_j$ of a few degrees on the scales of our simulations, for example 
$\theta_j = \SI{2}{\degree}$ out to $\SI{800}{au}$ in the atomic jet of RW Aur \citep{Dougados2000}
and $\theta_j \simeq \SI{5}{\degree}$ from $\SI{1000}{}-\SI{12000}{au}$ in the CO jet of IRAS04166+2706 \citep{Tafalla2009}. We explore a similar range in our simulations.

\subsection{Setup: Jet spray nozzle, radiative cooling, and AMR}
\label{section:monovariated_run:setup}

Similarly to \cite{Volker99}, we introduce a spray angle in the jet inlet 
by taking a velocity vector that is radially diverging from a virtual point: 

\begin{equation}
(\orthoradius,z)=(0,-z_0)\,, ~\hbox{ with } z_0=\orthoradius_j/\tan{\theta_j\,,}    
\end{equation}

\noindent where 
$\orthoradius_j$ is the jet radius at $z=0$ and $\theta_j$ the jet semi-opening angle. 

In order to conserve mass-flux throughout the jet inlet, 
we set a jet density profile $\rho_j(t)$ that decreases as the inverse square distance to the above-mentioned virtual point 
and varies inversely in time with the velocity modulus, as
\begin{align}
\rho_j(t) 
& = 
{\rho_{j0}}\left(\frac{\orthoradius_j^2+z_0^2}{\orthoradius^2+(z+z_0)^2}\right)
\frac{v_{0}}{v_j(t)}\,,
\label{eq:rho_j_t}
\end{align}
where $\rho_{j0}$ is the jet "base density" at $z=0$, $R = R_j$, and $t = 0$.
The (constant) jet mass-flux is then given by
\begin{align}
\massflowrate&=2\pi \left(1-\cos{\theta_j}\right)\left(\orthoradius_j^2+z_0^2\right)\rho_{j0}v_0\, 
\label{eq:massflux},\\
&=2\pi \left(1-\cos{\theta_j}\right)\left(\orthoradius_j/\sin\theta_j\right)^2 \rho_{j0}v_0.\,
\end{align}

Henceforth, the radiative cooling function at high temperatures ($T\geq 10^4~\mathrm{K}$) 
is taken from \cite{Schure09}, who take into account more up-to-date atomic parameters than 
\cite{MacDonald81}, whose radiative cooling function is applied in \cite{LeeEtAl2001} and included in \Sectref{section:PCJ}. 
Furthermore, the resolution and accuracy are improved by enabling AMR up to level 5, with a $\SI{84}{au}$ minimal and $\SI{5.2}{au}$ maximal spatial resolution. The computational grid is also expanded to $\SI{1.9e17}{\cm}$ in $z$.

\subsection{Model parameters}
\label{section:monovariated_run:parameters}

\begin{figure*}[!tp]
\ContinuedFloat*
    \centering
    \includegraphics[trim= 0.0cm 1.0cm 1.0cm 0.0cm, clip,scale=0.45]{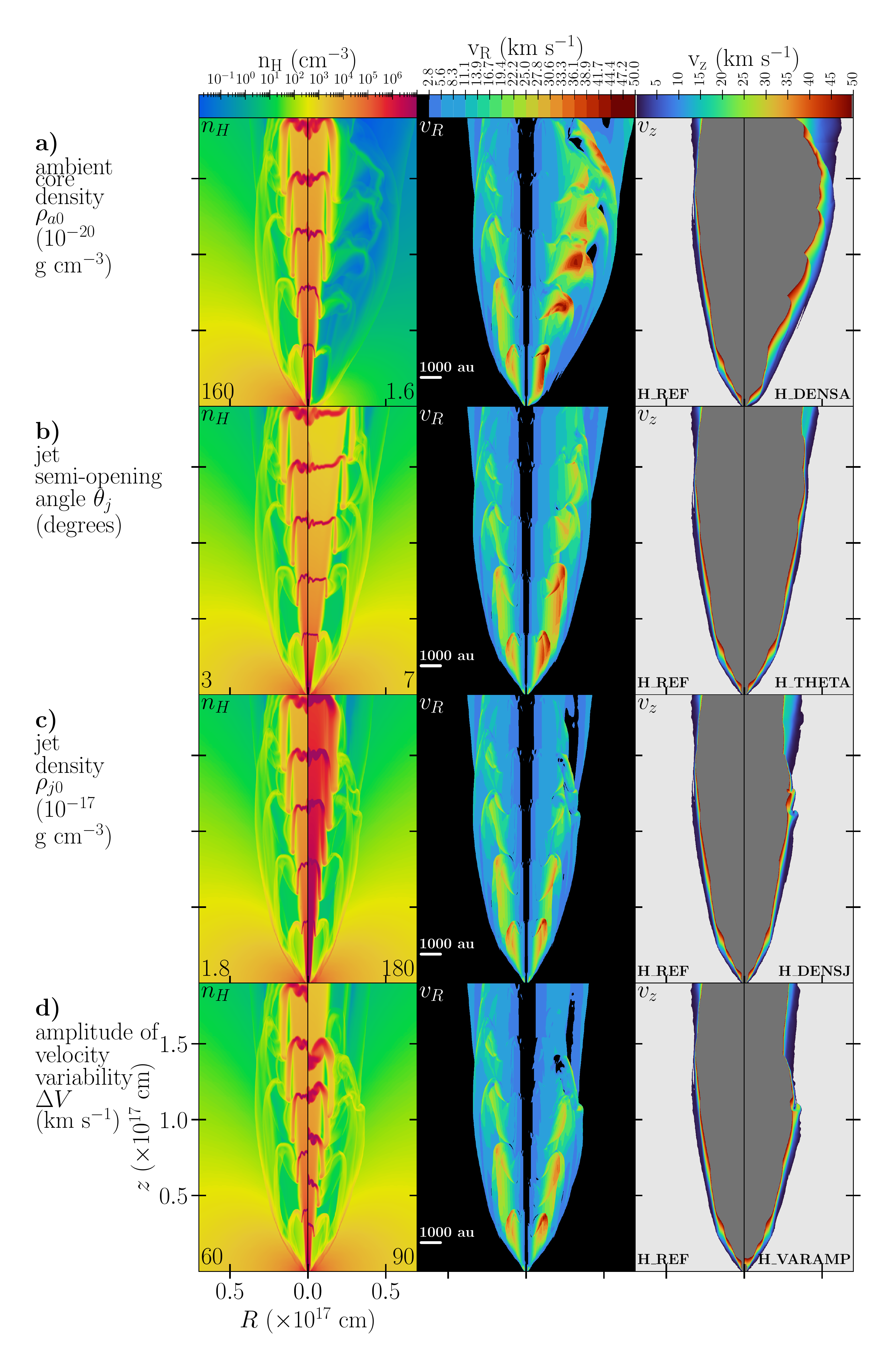}
    \caption{Influence of seven free parameters on the geometry and kinematics of shells driven by a pulsed conical jet in a singular flattened core, at an age of $t=\SI{700}{\yrs}$. Each row corresponds to a different free parameter (as labeled in the left margin) and compares maps of number density, $n_H$ (left), radial velocity, $v_\orthoradius$ (middle), and axial velocity, $v_z$ (right), for the reference model (left half of each map) and the modified model (right half of each map). The corresponding modified parameter values are marked at the bottom of the $n_H$ panel and model names at the bottom of the $v_z$ panel. Velocity colorbars are cropped to $0-\SI{50}{\kmps}$ for better visualization of the range detected in CO outflows. The reference model in this figure has a high mass-flux of $\SI{6e-6}{\MSun/yr}$ (see Table \ref{table:monovariated_run:parameters} for full list of model parameters). The main effect on the cavity shape is seen when varying the core density.
    }
    \label{fig:monovariated_run:maps}
\end{figure*}

\begin{figure*}[!tp]
\ContinuedFloat
\centering
    \includegraphics[trim= 0.0cm 1.0cm 1.0cm 0, clip,scale=0.435]{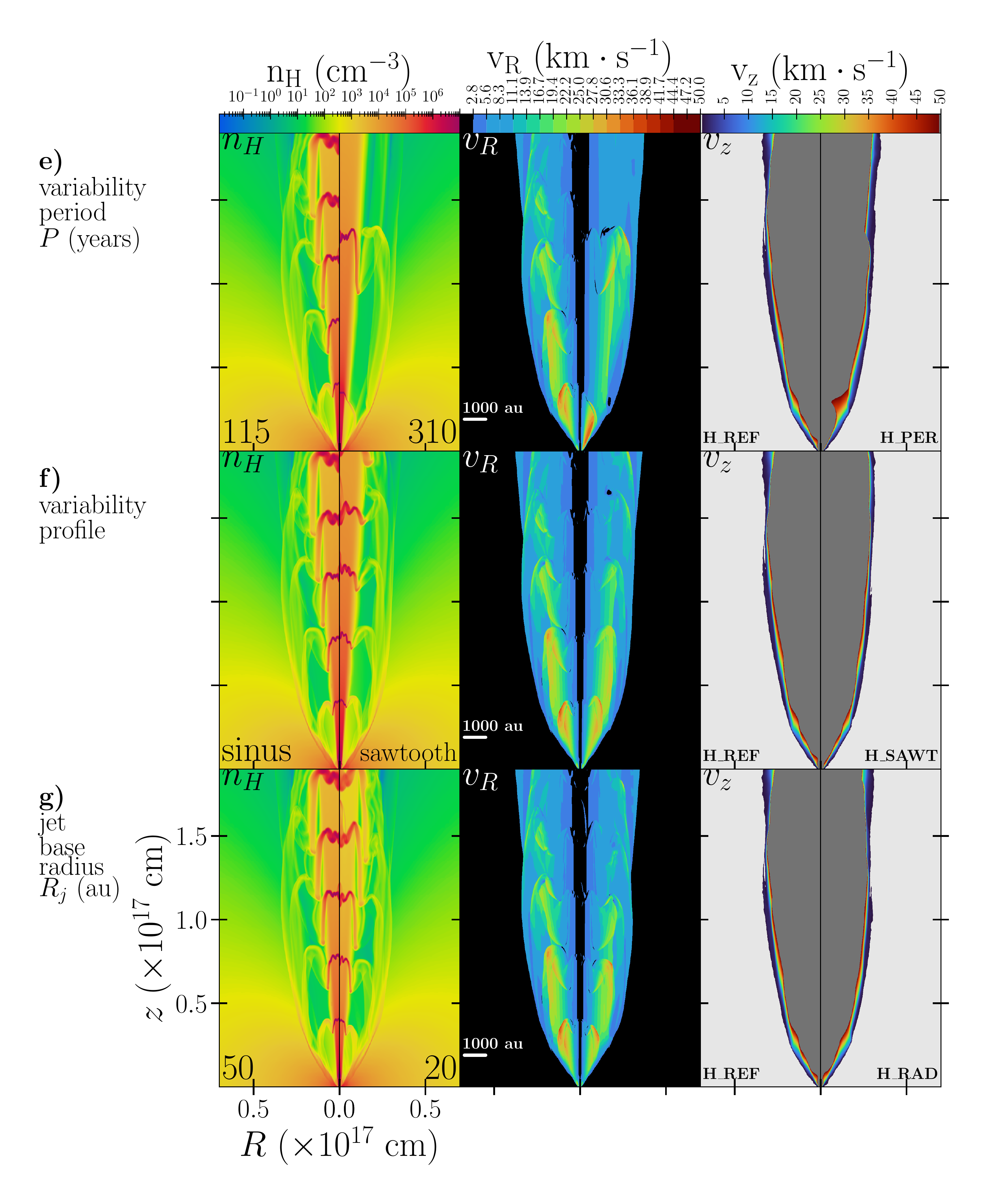}
    \caption{continued.} 
    \label{fig:monovariated_run:maps}   
\end{figure*}

  \begin{figure}[h!]
    \centering
    \resizebox{\hsize}{!}{\includegraphics[trim = 0 0.6cm 0 1.5cm ,clip]{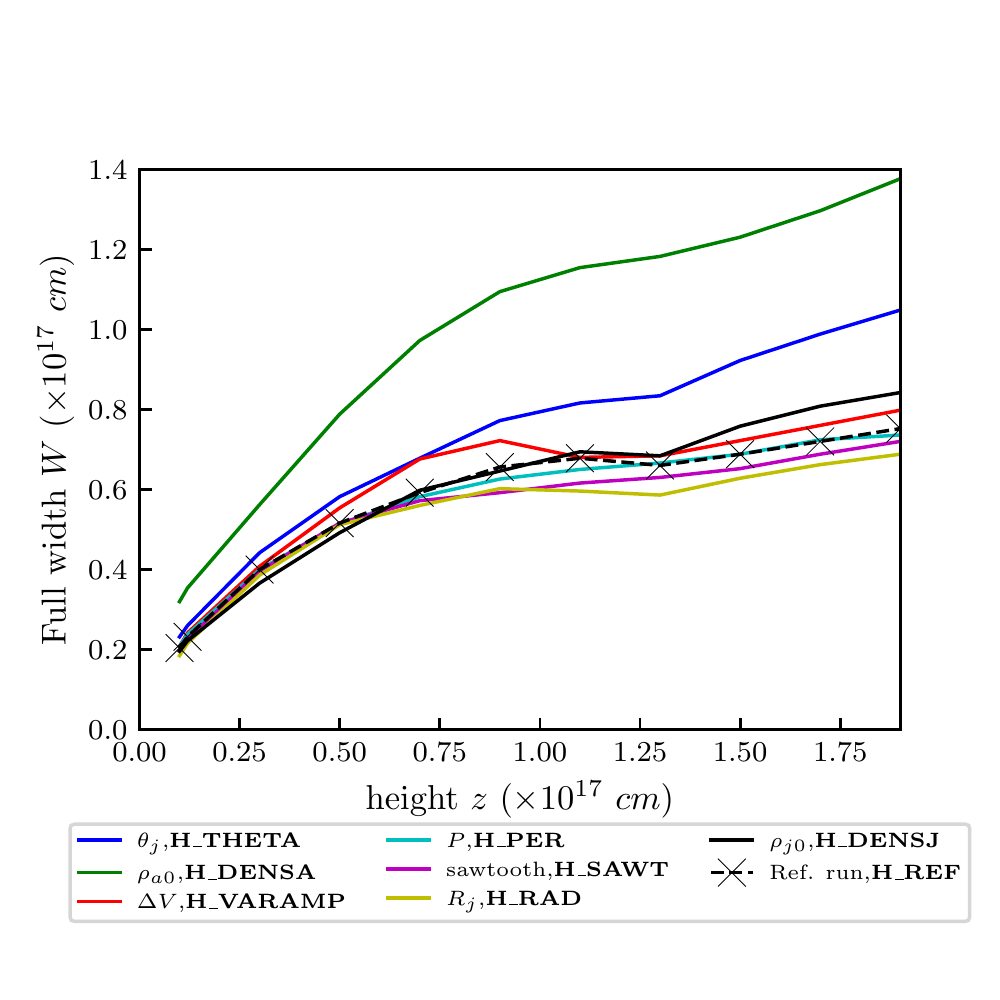}}
    \caption{Full shell width $\OpeningAngleWidth{}(z)$ at $t=\SI{700}{\yrs}$ as a function of altitude, $z,$ for the high-density jet models in \figref{fig:monovariated_run:maps}. Colored curves have one parameter varied from the reference run, from among the ambient core base density $\rho_{a0}$, jet semi-opening angle $\theta_{j}$, initial jet base density $\rho_{j0}$, semi-amplitude of variability $\Delta V$, period $P$, variability profile (sawtooth instead of sinusoidal), or jet radius $\orthoradius_j$. Model parameter values are listed in \Tableref{table:monovariated_run:parameters}. The covered measurements range from $z\sim\SI{800}{au}$ to $z\sim\SI{12700}{au}$.}
    \label{fig:monovariated_run:full_width}
\end{figure}

First, we defined a set of typical parameters for the reference models. For consistency, we kept the same ejection velocity law and ambient density distribution (singular flattened core) as in 
our model PCJ-RW in Table \ref{table:PCJ:parameters}. 
We adopted a reference value $\theta_j$ = 3\degr for the jet semi-opening angle, and 
a reference jet radius $\orthoradius_j =$ 50 au at $z=0$.

As for the jet base density $\rho_{j0}$, we considered a "high-density" value, 
giving a one-sided jet mass-flux $\massflowrate = \SI{6e-6}{\MSun/yr}$ typical of very active Class 0 protostars, for our reference values of $\theta_j$ and $\orthoradius_j$.

We then launched seven modified models, each having only one parameter changed with respect to the reference model (H\_REF): 

\textit{(i)}~H\_DENSA: with an ambient density scaling $\rho_{a,0}=\SI{1.6e-20}{\gram\per\centi\meter\cubed}$ instead of $\SI{1.6e-18}{\gram\per\centi\meter\cubed}$; 

\textit{(ii)}~H\_THETA: with a jet semi-opening angle $\theta_{j}=\SI{7}{\degree}$ instead of $\SI{3}{\degree}$; 

\textit{(iii)}~H\_DENSJ: with a jet base density $\rho_{j0}$ 100 times larger than the reference model; 

\textit{(iv)}~H\_VARAMP: with a variability semi-amplitude $\Delta V=\SI{90}{\kmps}$ instead of $\SI{60}{\kmps}$; 

\textit{(v)}~H\_PER: with a  variability period $P$ = $\SI{300}{\yrs}$ instead of $\SI{115}{\yrs}$; 

\textit{(vi)}~H\_SAWT: with a sawtooth velocity variability profile instead of a sinusoidal one; 

\textit{(vii)}~H\_RAD: with an initial jet radius $\orthoradius_j=\SI{20}{au}$ instead of $\SI{50}{au}$.

We note that all the models have the same value of jet mass-flux as the reference model H\_REF, except for H\_THETA, H\_DENSJ, and H\_RAD, since $\massflowrate$ independently varies with $\theta_j$, $\rho_{j0}$, and $R_j$ according to  \Eqref{eq:massflux}. 
All input parameters for the reference and modified models in this "high-density" case are summarized in \Tableref{table:monovariated_run:parameters}. 

To check the robustness of our conclusions, 
we also computed a second sequence of models with the same parameter changes, but with a 
100 times smaller reference jet density.
This leads to $\massflowrate = \SI{6e-8}{\MSun/yr}$ in the corresponding reference model (M\_REF) that is typical of more evolved Class 1 jets.
The parameters of this second sequence of models (referred to as the "medium-density" case)  are summarized in \Tableref{table:monovariated_run:parameters_B} and the results are summarized graphically 
in \figref{fig:appendix:full_width}.

\subsection{Results}
\label{section:monovariated_run:global_results}

\Figureref{fig:monovariated_run:maps} shows the differences in shape and kinematics of the shells between the reference and modified models in the high-density case, by comparing their respective maps of $n_H$ (H nucleus number density), $v_\orthoradius$, and $v_z$ at the same age of $t = \SI{700}{\yrs}$.
The $v_\orthoradius$ maps show the lateral expansion induced by successive bowshocks,
while the $v_z$ maps highlight the shear-like velocity gradient that develops along the main shell walls.

For an easier comparison between models, we use the density maps in \figref{fig:monovariated_run:maps} to measure the full width $\OpeningAngleWidth{}(z)$ of the main shell as a function of altitude $z$. We thus obtain \figref{fig:monovariated_run:full_width} that overplots the resulting shell shapes of each model. Finally, from the shell width $\OpeningAngleWidth{\SI{800}{}}$ at an altitude $z=\SI{800}{au}$, we derive the full opening angle near the base, defined following \cite{Dutta2020} as $\alpha_{800} = 2\arctan(\OpeningAngleWidth{800}/[2\times \SI{800}{au}])$.
Values of $\alpha_{800}$, $\OpeningAngleWidth{\SI{800}{}}$, and $\OpeningAngleWidth{\SI{12700}{}}$ (the full width at the top of the computational box) are summarized in the last three columns of \Tableref{table:monovariated_run:parameters}. 

The most impactful effect on both the morphology and kinematics of the main shell is obtained here when decreasing the core base density, $\rho_{a0}$, by a factor 100. In this case, \figref{fig:monovariated_run:maps}a and \figref{fig:monovariated_run:full_width} clearly show that
    the leading shell opens twice wider; $v_\orthoradius$ reaches higher values, which cover broader areas within the nested shells. The layer of strong $v_z$-gradient (between $0$ and $\SI{50}{\kmps}$) along the main shell surface also becomes thicker. 
    
    Increasing the jet opening angle (\Figsref{fig:monovariated_run:maps}b and \ref{fig:monovariated_run:full_width}) has a more moderate effect, with a maximum $\sim 30\%$ increase  of the main shell full width in comparison with the reference model in the high-density case. 
    
    The jet base density also has a moderate influence. 
    Decreasing the jet density by a factor 100 
    decreases the width of the shell at $z=\SI{800}{au}$ by at most $\SI{40}{\percent}$
    (comparing the curves of same color between \figref{fig:appendix:full_width} and \figref{fig:monovariated_run:full_width}).
    
    Finally, the shape and kinematics of the main shell are little affected when semi-amplitude $\Delta V$ (\figref{fig:monovariated_run:maps}d), variability period $P$ (\figref{fig:monovariated_run:maps}e) are increased, when a sawtooth profile $h(t)$ is superimposed (\figref{fig:monovariated_run:maps}f), or when the jet radius $\orthoradius_j$ (\figref{fig:monovariated_run:maps}g) is decreased. 
    
    We obtain the same behaviors for a  jet density that is 100 times smaller
    (see \figref{fig:appendix:full_width}).

However, we find that in comparison with the sinusoidal case, a sawtooth-like profile of variability leads to smoother shells borders and less unstable bowshocks and IWS, as well as reduced instabilities in general (cf. \figref{fig:monovariated_run:maps}f).
    
\section{Millenia-long simulations of a conical pulsed jet in a stratified core}
\label{section:long_term}

In this section, we investigate the long-term evolution of the shells
driven by a pulsed conical jet through a flattened singular core. To do so, for the first time we present this type of simulation reaching up to $\SI{10000}{\yrs}$. Predicted sizes, position-velocity (PV) diagrams, and mass-velocity distributions are presented and qualitatively compared with the typical behavior in Atacama Large Millimeter/sub-millimeter Array (ALMA) observations of outflows.

\subsection{Setup}
\label{section:long_term:setup}

In \Sectionref{section:monovariated_run}, we
noted that a sawtooth jet velocity variability profile minimizes the development of instabilities without changing the overall shell shape and kinematics. We thus adopted such a sawtooth profile here, since it allows us to reach the desired long timescales at a more reasonable CPU cost.
For consistency, we adopted the same jet and ambient parameters as in model M\_SAWT in Table A.1, except for the variability period. We considered here a slightly longer value $P=\SI{300}{\yrs}$, as inferred from CO observations of both the HH46-47 outflow by \cite{Zhang19}
and the CARMA-7 outflow by \cite{Plunkett15}, before the inclination correction. 

To keep a non-prohibitive computational time, we also use here a more diffusing Total Variation Diminishing Lax Friedrichs numerical scheme (TVDLF) at the shocks zone.  This decreases the instabilities at the highest resolution, without changing the overall shell structure.
Finally, to follow the outflow expansion, the computational domain is expanded to  $\SI{3.0e17}{\cm}=\SI{20000}{au}$ in $\orthoradius$ and $z$ ($n_\orthoradius\times n_z =240\times240$ cells at AMR level 1), keeping the same resolution as in \Sectref{section:monovariated_run} and Table 2.

\subsection{Long-time maps of density, mixing fraction, and velocities}
\label{section:long_term:maps}

\begin{figure*}[!t]
    \centering
    \includegraphics[trim= 4.0cm 1.7cm 3.0cm 1.0cm, clip,width=\textwidth]{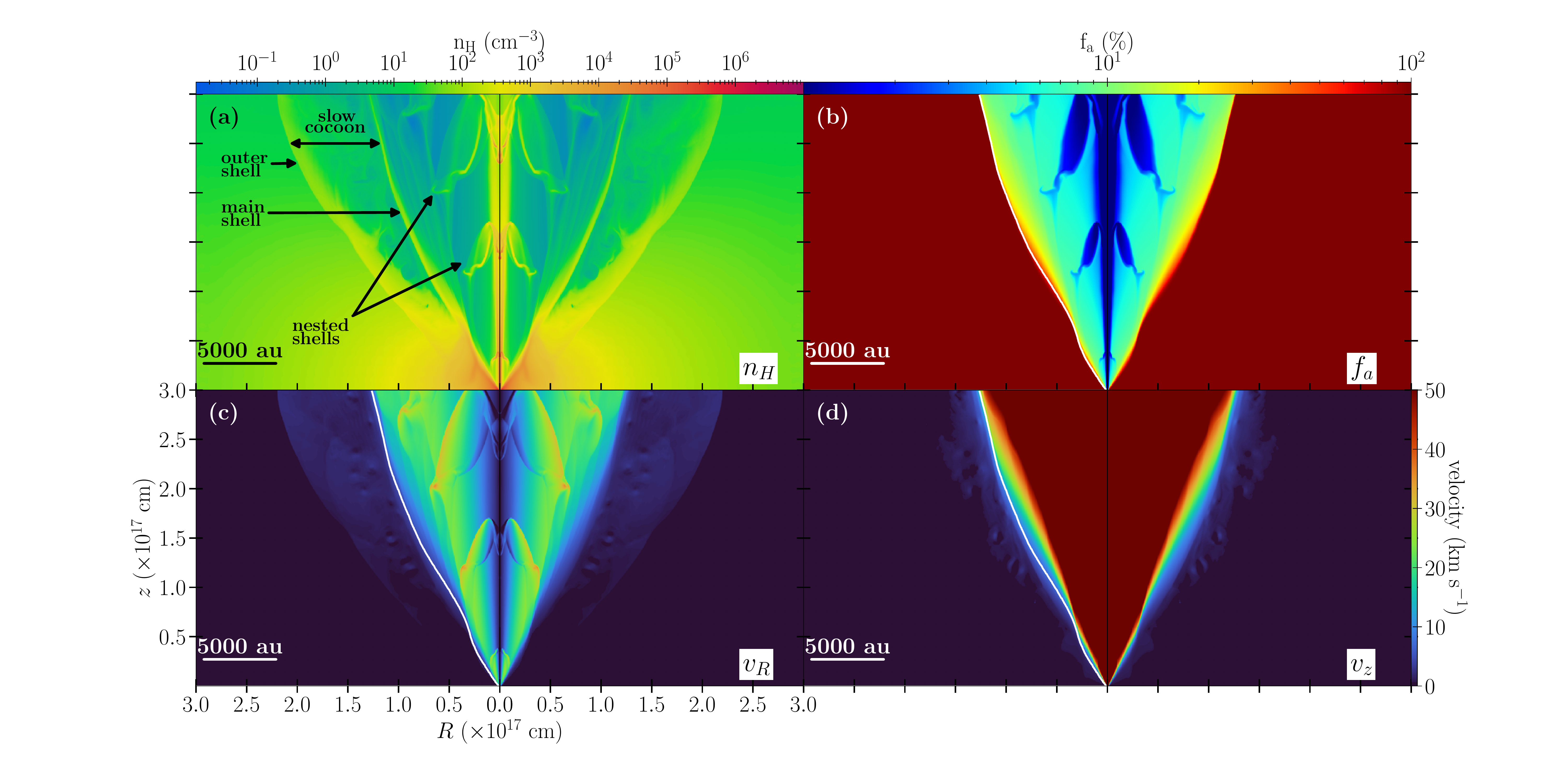}
        \caption{Maps at $t=\SI{10000}{\yrs}$ of (a) hydrogen nuclei density $n_H$, (b) fraction of core-originated material $f_a$, (c) orthoradial velocity $v_R$, and (d) axial velocity $v_z$ of the shell driven by a conical pulsed jet in a flattened singular core. Parameters are identical to model M\_SAWT in \Tableref{table:monovariated_run:parameters_B} except for a longer variability period $P=\SI{300}{\yrs}$.
        Velocities in the colorbars are cropped to the range typically detected in CO outflow observations, namely between $\SI{0}{}$ and $\SI{50}{\kmps}$. The white contour in the left-half of panels of (b), (c), and (d) shows $f_a=50~\%$.}
    \label{fig:long_term:maps}
\end{figure*}

\Figureref{fig:long_term:maps} shows the distribution of number density, $n_H$, 
fraction of core-originated material, $f_a$, and 
orthoradial and axial velocities, $v_R$ and $v_z$, in
the outflow at $t=\SI{10000}{\yrs}$.
\Figureref{fig:long_term:maps}a shows that 
after ten thousand years, the outflow structure differs from that seen at earlier ages (e.g., in  \figref{fig:monovariated_run:maps}). The initially single shell has split into two separate shells: an "outer shell" tracing the forward shock propagating at low speed into the ambient medium and an inner "main shell" tracing the jet-ambient interface (roughly delimited by the $f_a=\SI{50}{\percent}$ border), where the wings of successive bowshocks pile-up. Both shells display a roughly parabolic or conical shape out to $\SI{20000}{au}$. 
Between these two shells is a slow cocoon of ambient material. Inside the main shell, we can discern several distinct "nested shells,"  tracing the last bowshocks recently created by jet variability. This geometry is reminiscent of the "spider-like" structure observed at the base of the B5-IRS1 outflow, where an inner parabolic shell, with a jet shock at its apex, is nested inside the wide-angle low-velocity outflow cavity \citep{Zapata2014}.

Moreover, the mixing map (\figref{fig:long_term:maps}b) shows that a fraction of material from the surrounding core can go past the shock and mix with jet-originated material inside the 
main cavity up to the wings of the nested bowshocks. Some core-originated material can even reach particularly overdense areas bordering the jet walls, with $n_H$ between $10^2$ and $10^5~\mathrm{cm^{-3}}$. In the following, we assume that this core-originated material remains molecular for the purposes of computing the predicted synthetic emission diagrams.

Finally, \figref{fig:long_term:maps}c and \ref{fig:long_term:maps}d show the radial and axial velocity maps, respectively. The new bowshocks encounter less resistance than those at earlier ages and develop broader wings, as the outflow cavity has been cleared up by tens of older bowshocks. Nevertheless, the characteristic kinematic pattern remains similar to early times, with enhanced-$v_R$ in the bowshocks wings and a shear-like gradient of axial velocity $v_z$ along the main shell at the jet-ambient interface. This velocity structure creates a characteristic "bell-shaped" pattern in transverse position-velocity diagrams, presented in Sect. \ref{section:long_term:PV_transverse}.

\subsection{Deceleration of the jet-driven shells}
\label{section:long_term:time_full_width}

In contrast to a wind-driven shell in a flattened singular core, which expands at constant speed over time \citep{Shu91}, a jet-driven shell is expected to decelerate. However, no analytical estimate of that deceleration exists in the case of a non-uniform ambient medium, making numerical simulations necessary.

\begin{figure}[t!]
    \resizebox{0.93\hsize}{!}{\includegraphics[trim = 0.3cm 0.3cm 0.3cm 0.1cm,clip]{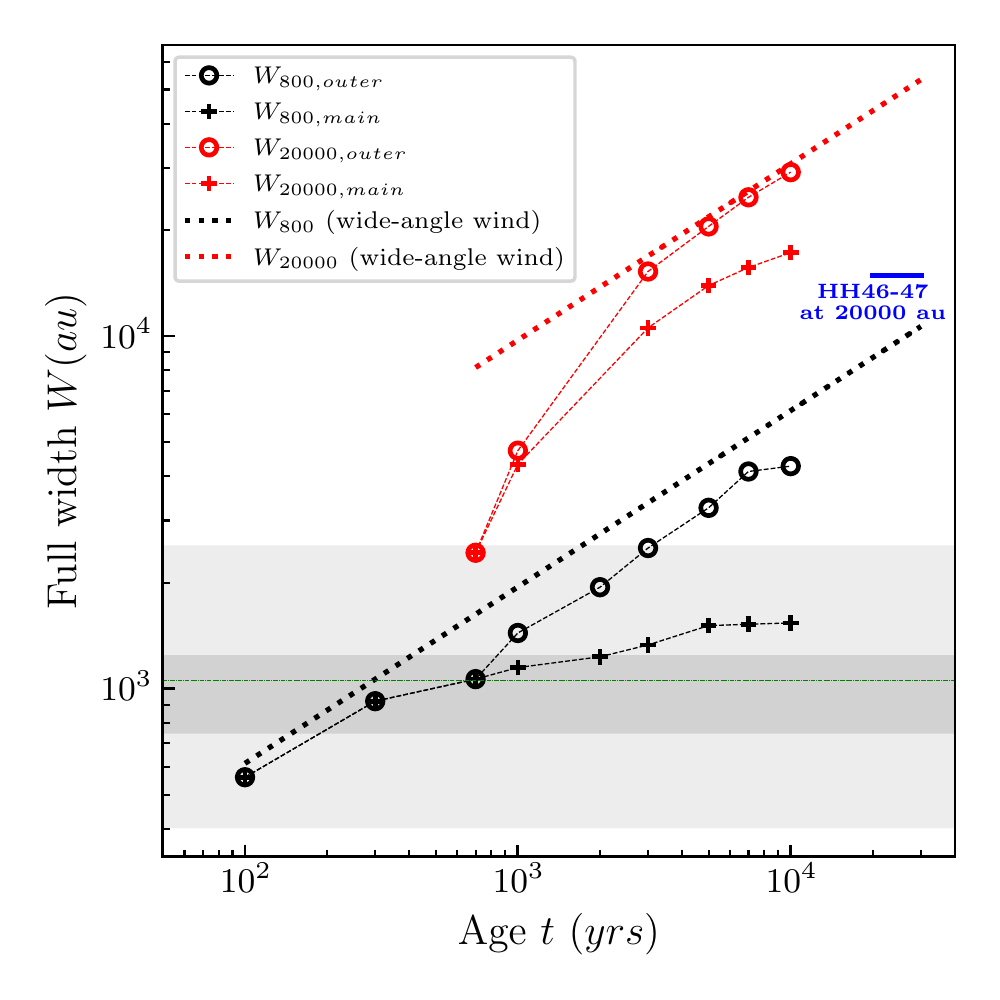}}
        \caption{Temporal evolution in the simulation of \figref{fig:long_term:maps} of the full width of the main shell (jet-ambient interface, crosses) and outer shell (forward shock, circles) measured at altitudes $z=\SI{800}{au}$ (black) and $z=\SI{20000}{au}$ (red). Analytical predictions for the equivalent "modified X-wind"
        model of \cite{LeeEtAl2001} are shown as dotted black and red lines of slope $t^{1/2}$ (see text). 
        The full range of CO outflow widths observed at $z= \SI{800}{au}$ by \cite{Dutta2020} is indicated by the
        light grey band (with second and third quartiles in darker grey, and median as a thin green line). The full width of the HH46-47 outflow at $z = \SI{20000}{au}$, from \cite{Zhang16}, is shown in {blue}.}
    \label{fig:long_term:time_full_width}
\end{figure}

\Figureref{fig:long_term:time_full_width} shows the time evolution of the shell widths during the $\SI{10000}{\yrs}$-long simulation
 leading to the snapshot of \figref{fig:long_term:maps}. We plot the full width as function of time at two different heights:  $z=\SI{800}{au}$
\citep[for comparison with][]{Dutta2020}, 
and $z=\SI{20000}{au}$ (top of the computational domain, after the jet head reaches it). We denote the corresponding shell widths as 
$W_{800}$ and $W_{20000}$.

\Figureref{fig:long_term:time_full_width} shows that the formation of two separate shells 
(denoted as the outer and main shells in \figref{fig:long_term:maps}) occurs around $t=\SI{1000}{\yrs}$. 
Both shells are seen to decelerate at late times. 
Deceleration is stronger at lower altitudes, where the ambient core is denser. At $z = \SI{800}{au}$,  the main shell stops expanding after $\SI{6000}{\yrs}$ and reaches a final width $W_{800} \simeq \SI{2e16}{\cm} = \SI{1500}{au}$. 
The corresponding final opening angle is $\alpha_{800} = 86\degr$.
This behavior is consistent with observations suggesting that the base opening angle of CO outflows stops increasing after $t \simeq$ 8000 yr, with a final value (uncorrected for inclination) $\alpha_{\rm obs} \simeq 90-100\degr$ \citep{Velusamy2014}.

The widths $W_{800}$ of the main shell at $z=\SI{800}{au}$ 
also fit very well within the observed range of flow widths 
at the same projected height 
(indicated by grey bands in \figref{fig:long_term:time_full_width}), 
measured  by \cite{Dutta2020} in a sample of 22 CO outflows in Orion.
On large core scales of $z = \SI{20000}{au}$ = 0.1 pc, the main shell reaches a width $W_{20000} = \SI{15000}{au}$ at an age of $10^4~\SI{}{\yrs}$. This is similar to the observed CO outflow width at the same (deprojected) height in HH46-47 \citep{Zhang16}, indicated in blue in \figref{fig:long_term:time_full_width}. On intermediate scales of $z = \SI{8000}{au}$, the main shell width at an age of $10^4~\SI{}{\yrs}$ in our simulation is 9000 au (see Fig. \ref{fig:long_term:maps}). This is also in good agreement with cavity widths observed at the same (projected) distance in scattered light, lying in the range 1100--8500 au in 75\% of cases \citep[cf. semi-opening angles reported in][]{Habel2021}.
Therefore, a jet driven into a flattened singular core seems able to reproduce typical observed outflow widths on both small and large scales for realistic long ages of $\ge \SI{10000} {\yrs}$.

For comparison purposes, the dotted lines in \figref{fig:long_term:time_full_width}  show the predicted evolution of shell widths at $z =$ 800 and 20000 au for the equivalent wide-angle "modified X-wind" model of \cite{LeeEtAl2001}, which has the same mean axial velocity and total mass-flux as our jet and propagates in the same flattened singular core.
In this model, the wind velocity drops away from the axis as $V_w = V_{w0} \cos\theta$, and the base of the shell is a parabola, $z = R^2/R_0$, expanding self-similarly over time as $R_0 = V_0 t$, where $V_0 = V_{w0} \sqrt{\eta}$ with $\eta$ the 
(fixed) ratio of wind-to-ambient density in the equator. 
The shell width at any fixed height $z$ is then given by $W_z = 2R = 2 \sqrt{z V_0 t}$. 
In the equivalent wide-angle wind model considered here, $V_{w0} = \SI{120}{\kmps}$ and $\eta = 2.2 \times 10^{-3}$ \citep{LeeEtAl2001}, yielding $V_0 = \SI{5.6}{\kmps}$. 
Observed outflow shells fitted with the same analytical model\footnote{
\cite{Lee2000} use different notations, $C = 1/R_0$ and $v_0 =1/t$} have similar values of $V_0 \simeq 1.5-9~\SI{}{\kmps}$  \citep{Lee2000,Zhang19}. 

The black dotted line in \Figureref{fig:long_term:time_full_width} shows that while the wide-angle wind-driven shell predicts comparable widths
to the jet-driven shell at early times $\le 300$ yr (as noted in Section 3), its self-similar expansion without deceleration
exceeds observed CO outflow widths at $z = 800$ au (grey band in \Figureref{fig:long_term:time_full_width}) after only two thousand years, and the discrepancy increases over time as $t^{1/2}$.
Recent simulations of X-wind-driven shells
including magnetic fields in both the wind and the ambient medium predict similar shell widths at 500-1000 years as this simple analytical model
and confirm that the shell expands at constant rate in a self-similar fashion, $R \propto t$ \citep{Shang20}. We note that this ongoing expansion is tightly linked to the adopted $1/r^2$ ambient density distribution \citep{Shu91}. Steady wind-driven shells could form if the core has a shallower density gradient and if mixing is inefficient at the wind-core interface \citep{Smith86,Liang20}.

\subsection{Synthetic predictions}

\subsubsection{Longitudinal position-velocity diagrams}
\label{section:long_term:inclination_angle}

 \begin{figure*}[t!]
    \centering
    \resizebox{0.87\hsize}{!}{\includegraphics[trim= 2cm 1.75cm 3cm 1.35cm, clip]{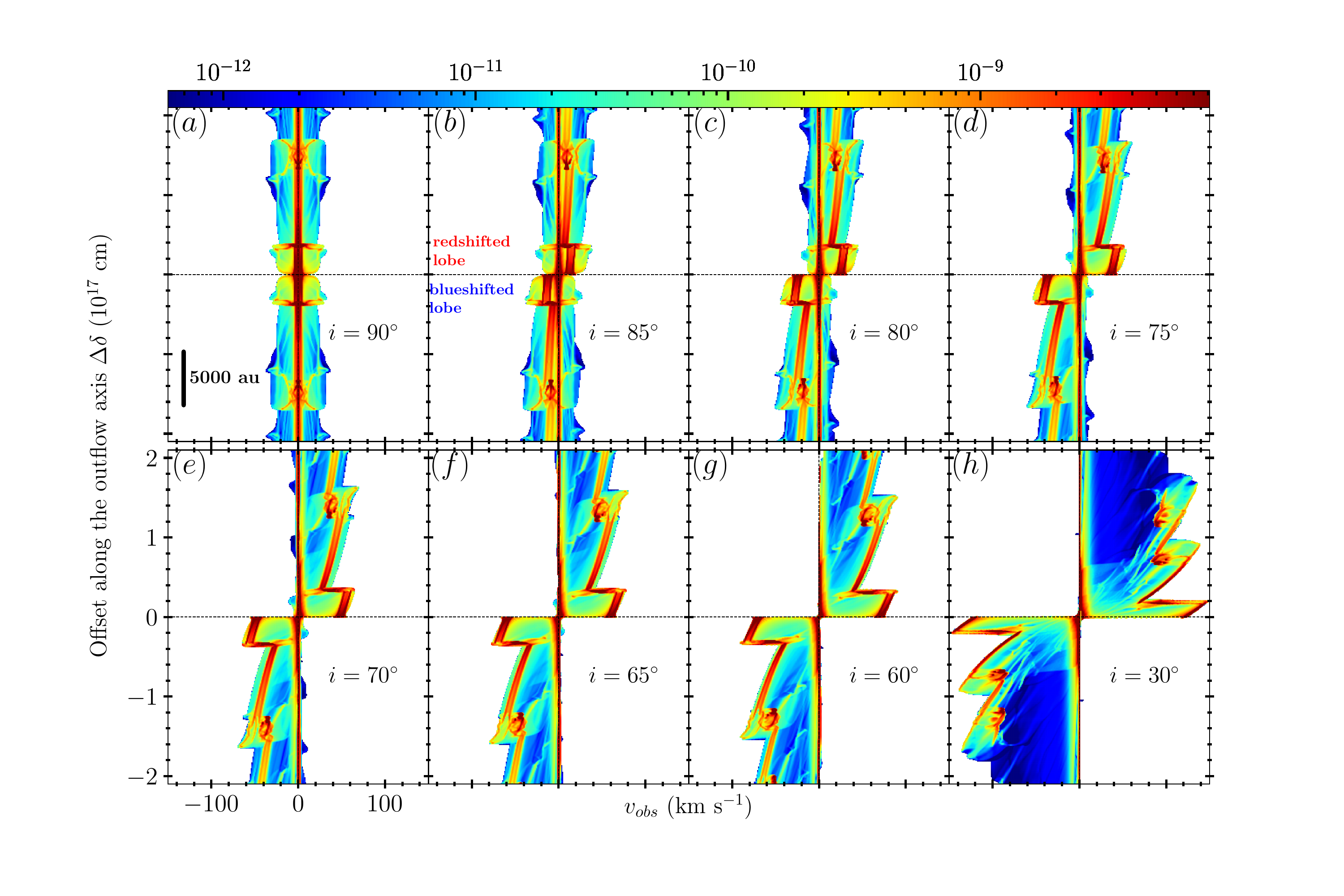}}
    \caption{Longitudinal position-velocity diagrams along the flow axis, inferred from the simulation in 
    \figref{fig:long_term:maps} at an age $t=\SI{10000}{\yrs}$, for different inclination angles, $i,$ from the line of sight, ranging from $\SI{90}{\degree}$ (edge-on) to $\SI{30}{\degree}$.} 
    \label{fig:long_term:PV_inclination_angle}
\end{figure*}

\begin{figure*}[h!]
    \centering
    \resizebox{0.7\hsize}{!}{\includegraphics[trim= 6.5cm 2cm 7.0cm 0, clip]{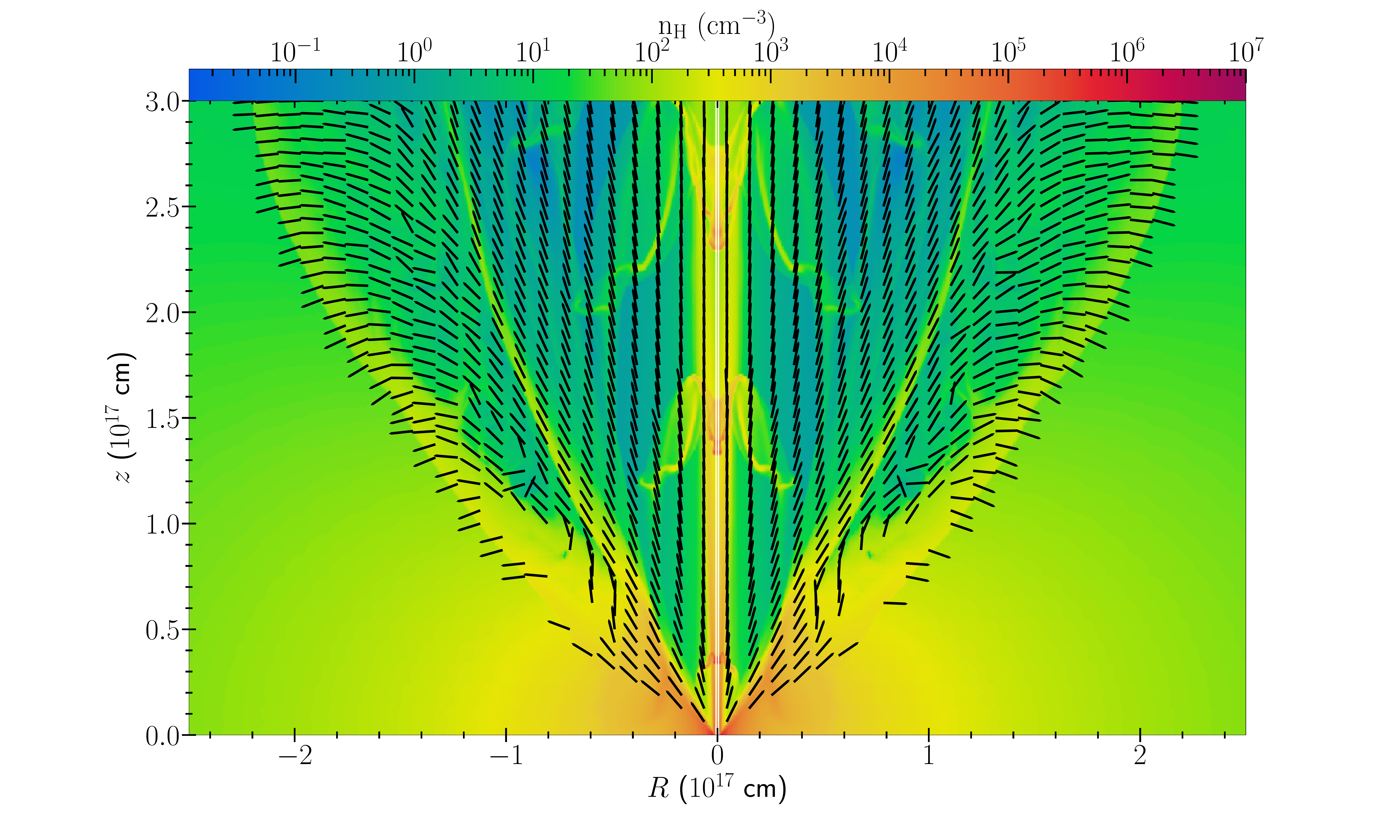}}
    \caption{Orientation of the velocity vectors for the $t=\SI{10000}{\yrs}$-long simulation in \figref{fig:long_term:maps}, in the reference frame of the ambient core. Arrows are shown for velocities above \mialyadd{$\SI{0.45}{\kmps}$}, and are not scaled with the velocity modulus. Density is displayed in the background color image.} 
    \label{fig:long_term:arrows}
\end{figure*}

\Figureref{fig:long_term:PV_inclination_angle} presents longitudinal position-velocity (PV) diagrams cut along the jet axis from our long-term simulation at $t=\SI{10000}{\yrs}$, 
assuming various inclination angles, $i,$ of the blueshifted lobe from the line of sight.

\begin{figure*}[h!]
    \centering
    \includegraphics[trim= 3.125cm 4cm 5cm 3cm, clip,width=\textwidth]{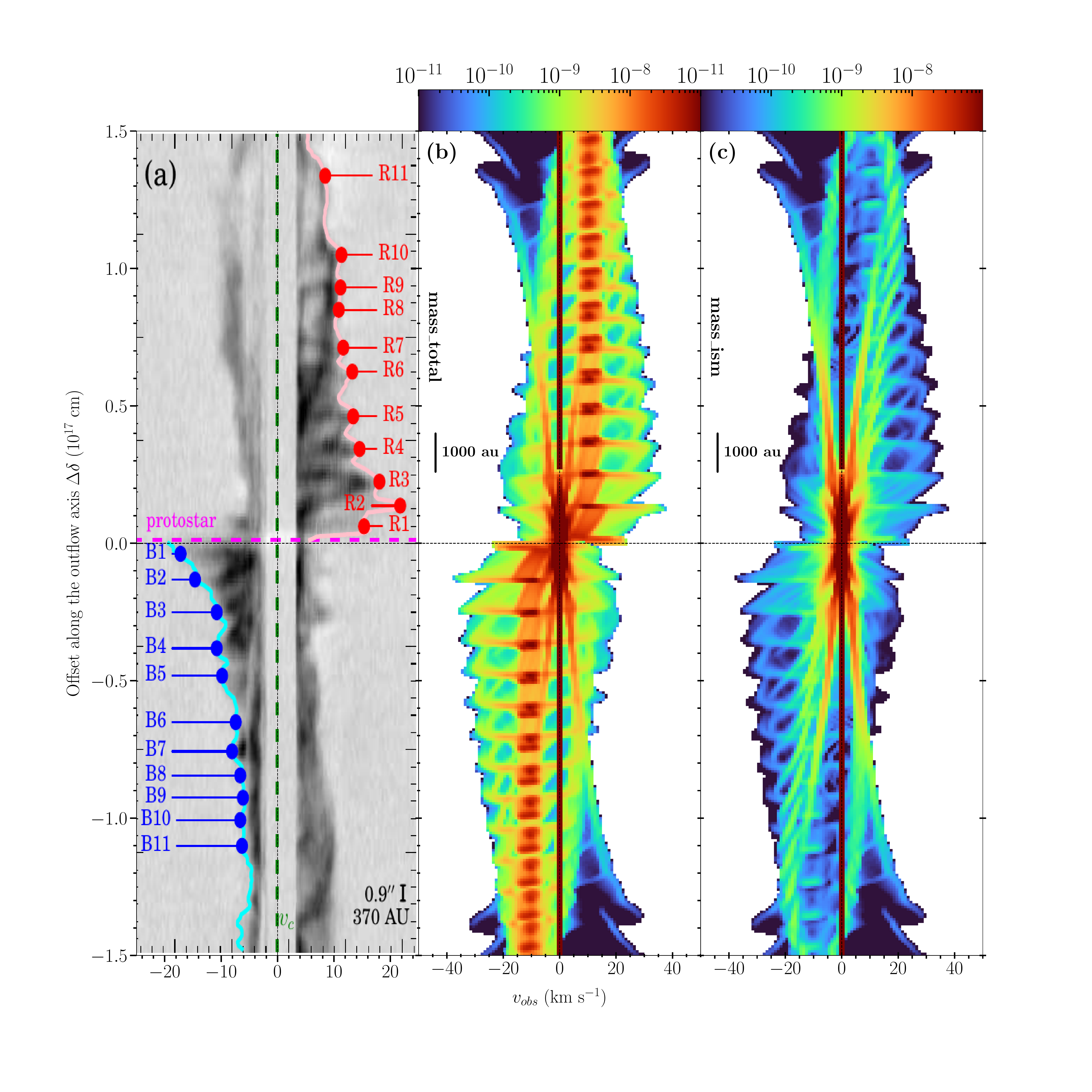}
    \caption{Longitudinal position-velocity (PV) diagrams. (a) Longitudinal PV diagram of CARMA-7 inside a $\SI{374}{au}$-wide cut along the outflow axis, as observed in $\ce{{}^12CO(2-1)}$ by \cite{Plunkett15}. \textit{(b)} Synthetic longitudinal PV diagram 
    along a $\SI{400}{au}$-wide cut\mialyadd{, at $t=\SI{500}{\yrs}$,} and inclination $i=\SI{85}{\degree}$ from the line of sight. \mialyadd{Parameters are the same as for the model shown in \figref{fig:long_term:maps} except for a shorter jet variability period $P=\SI{30}{\yrs}$.}  \textit{(c)} Same as \textit{(b)} but including only the core-originated material. }
    \label{fig:long_term:Plunkett2015}
\end{figure*}

Most importantly, this figure shows that 
at ages typical of Class 0 outflows, negative velocities all vanish below an inclination angle of  $i < \SI{70}{\degree}$ from the line of sight. In particular, there is no more observable blue/red overlap at $i = \SI{60}{\degree}$, unlike what has been predicted for jet-driven shells in uniform media \citep{LeeEtAl2001}. 

This behavior is induced by the transverse deceleration of the main shell on late timescales, as discussed in \Sectref{section:long_term:time_full_width}. The slow expansion of the main shell restricts the sideways expansion of the bowshock wings propagating inside it, forcing them to adopt more forward-directed velocities. This may be seen in \figref{fig:long_term:arrows}, where we  plot the direction of velocity vectors in the outflow. Along the main shell, where bowshock wings pile up, velocity vectors are almost parallel to the shell walls. This produces much less blue-red overlap in the projection than is otherwise expected for a bowshock in a uniform medium. In the outer shell, which traces the forward shock expanding into the ambient medium, the velocity vectors are widely open (perpendicular to the shell) but the local expansion velocity is so low that the emission falls near rest velocity once it is projected. 

Challenges to the notion of jet-driven shells producing too much blue-red overlap over a broad range of view angles thus no longer appears justified when realistic long ages and ambient core stratifications are considered.
A longitudinal PV cut extending up to $\SI{1.5e17}{\cm}$ from the source was recently obtained in the CARMA-7 outflow with ALMA \citep{Plunkett15}, with both high resolution and high  sensitivity.

A qualitative comparison with our predictions is presented in \figref{fig:long_term:Plunkett2015}. 
Since the PV of CARMA-7 presents significant blue-red overlap, it is believed to be close to the plane of the sky, hence, we considered an inclination of $i=\SI{85}{\degree}$ to the line of sight. 
The observed PV cut, reproduced in \figref{fig:long_term:Plunkett2015}a, shows a striking quasi-periodic series of velocity peaks. The differences in dynamical timescales between successive identified  
velocity peaks yield an apparent period of variability in CARMA-7 of $\Delta\tau_{\rm dyn} \simeq \SI{300}{\yrs}$  \citep[see][]{Plunkett15}. However, for a quasi edge-on inclination, our synthetic PV diagrams with $P = \SI{300}{\yrs}$ in \figref{fig:long_term:PV_inclination_angle} predict a much wider knot spacing than observed in \figref{fig:long_term:Plunkett2015}a. As noted by \cite{Plunkett15}, the value of  $\Delta\tau_{\rm dyn}$ may need to be corrected for projection effects by a typical factor 10. 
We thus 
present in \figref{fig:long_term:Plunkett2015} a model with
 jet variability period that is ten times shorter, namely, $P \simeq \SI{30}{\yrs}$.  

Two synthetic PV diagrams are presented in \figref{fig:long_term:Plunkett2015}: one including all material from jet and core (\figref{fig:long_term:Plunkett2015}b) and the other including only core-originated material (\figref{fig:long_term:Plunkett2015}c). 
 \Figureref{fig:long_term:Plunkett2015}b is dominated by periodic sawtooth structures tracing the time variable jet and the sideways ejected material from its internal working surfaces (IWS).  \Figureref{fig:long_term:Plunkett2015}c is dominated by structures with apparent "Hubble-law" acceleration, tracing ambient gas swept-up in the successive nested bowshock wings. We may remark that the jet and IWS remain faintly visible, through ambient gas dynamically entrained along the jet borders (cf. mixing map in \figref{fig:long_term:maps}b).

We find that the predicted structures in \figref{fig:long_term:Plunkett2015}c are qualitatively similar to what is observed in CO emission in CARMA-7 (\figref{fig:long_term:Plunkett2015}a). Since our model is very simplified (e.g., it does not include any jet precession), this qualitative agreement can be considered as promising. It also confirms that a suitable inclination-correction is essential to estimate the true period of velocity variations in a quasi edge-on outflow.

\begin{figure*}[t!]
    \centering
    \includegraphics[trim= 0 1.0cm 0 0, clip,width=\textwidth]{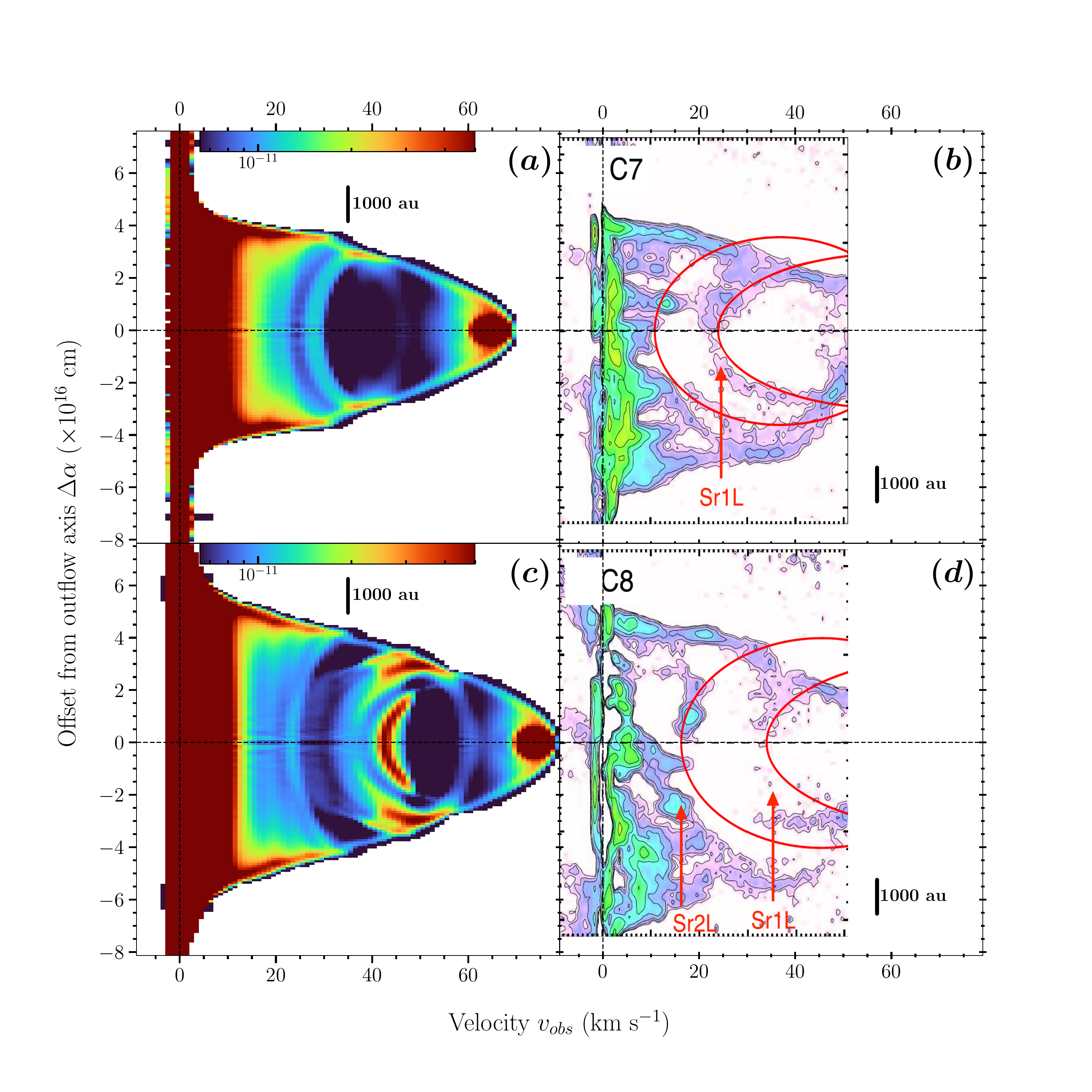}
    \caption{Transverse PV diagrams.  \textit{Left panels:} Transverse position-velocity diagrams perpendicular to the outflow axis, inferred from the jet-driven shells simulation in \figref{fig:long_term:maps} and \Sectref{section:long_term:maps}, at $t=\SI{10000}{\yrs}$ inside $\SI{400}{au}$-wide cuts at projected heights of (a) $\SI{6.6e16}{\cm}$ and (c)~$\SI{8.9e+16}{\cm}$, with an outflow inclined by an angle $i=\SI{55}{\degree}$ from the line of sight. Here only the core-originated material mass contribution is shown. {\textit{Right panels}:}  \ce{^{12}CO(2-1)} emission of HH46/47 along two $\SI{450}{au}$-wide cuts at projected heights of (b) $\SI{8.1e16}{\cm}$ and (d) $\SI{1.0e17}{\cm}$ from the central source, with red ellipses showing best-fit models by wide-angle wind-driven shells. Adapted from \citet{Zhang19}. 
    }
    \label{fig:long_term:PV_transverse}
\end{figure*}

\subsubsection{Transverse PV diagrams}
\label{section:long_term:PV_transverse}

\Figuresref{fig:long_term:PV_transverse}a,c show synthetic transverse PV diagrams for our $\SI{10000}{\yrs}$-old simulation of an outflow with $P=\SI{300}{\yrs}$ (same simulation as in \figref{fig:long_term:maps}). We adopted an inclination angle $i=\SI{55}{\degree}$ from the line of sight, the mean inclination of the HH46-47 jet determined by \cite{Hartigan05}.
First, each diagram forms a characteristic "bell-like" shape, which peaks at  high-velocity
(the jet) and broadens smoothly down to rest velocity. This shape is a direct consequence of the deceleration of bowshock wings as they expand and interact inside the shell, which produces a velocity decreasing away from the jet axis. Second, while most of the mass is piled-up near rest velocity, and in the $v_z$ "shear layer" along the shell walls (responsible for the two "horns" along the edges of the bell), bright ellipses are also present at intermediate velocities. Those ellipses trace intersections of the line of sight with individual bowshocks expanding inside the main shell.

These predicted characteristics 
(bell-shape with nested ellipses)
bear striking qualitative resemblance with observed transverse PV cuts \mialyadd{at high resolution and sensitivity}, recently obtained  with ALMA across the HH46-47 outflow by \cite{Zhang19} and shown for comparison in \figref{fig:long_term:PV_transverse}b,d. 
The agreement appears even better than with models of wide-angle wind-driven shells \citep{Zhang19}, which tend to overpredict the outflow width at high projected velocities (see red ellipses on top of the observed PVs in \Figsref{fig:long_term:PV_transverse}b,d).

\subsubsection{Mass-velocity distribution}
\label{section:long_term:mv}

\begin{figure}[t!]
    \centering
    \resizebox{\hsize}{!}{\includegraphics[trim= 0 1.0cm 0 0, clip]{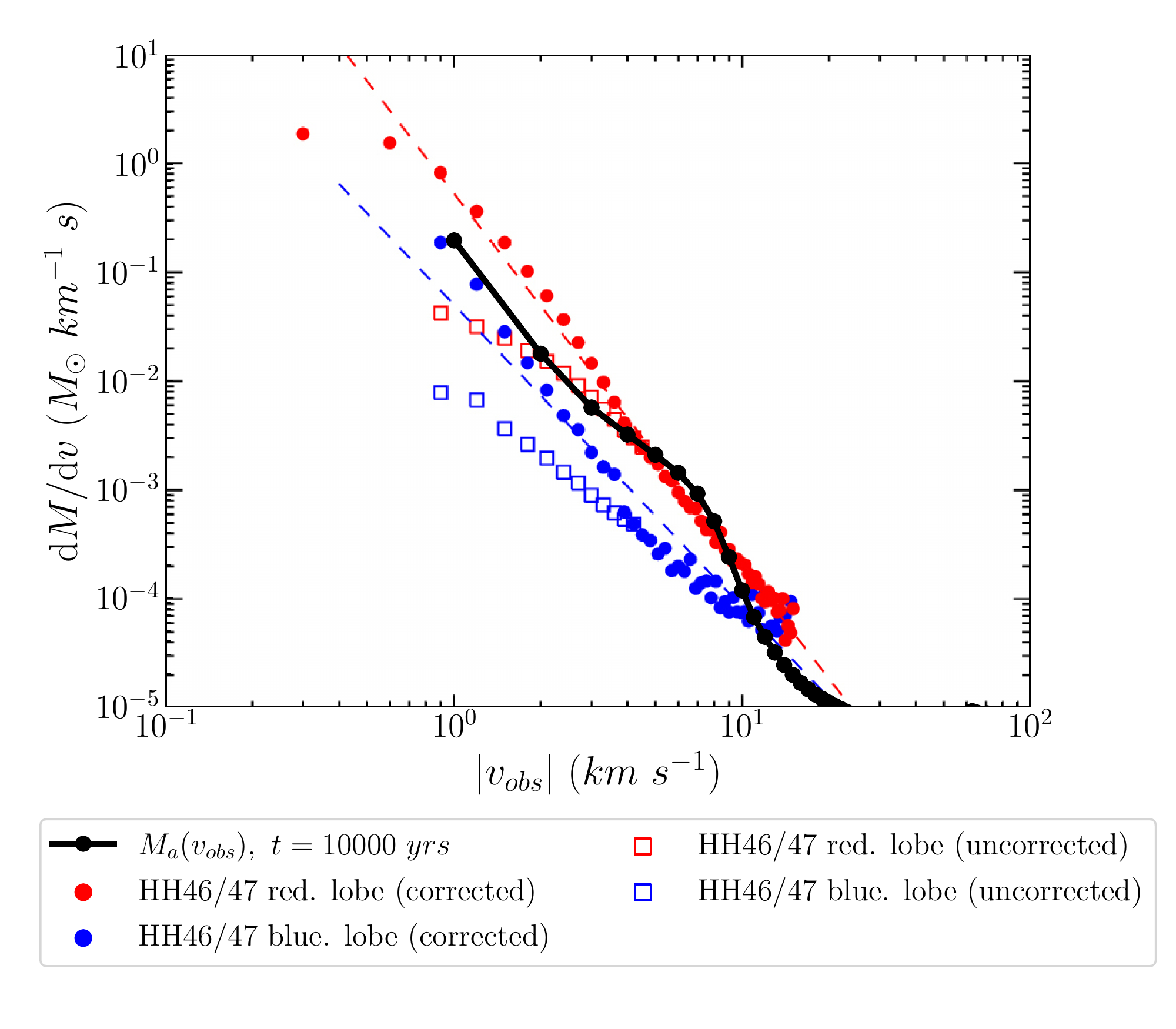}}
    \caption{Simulated mass-velocity (MV) distribution of core-originated material at $t=\SI{10000}{\yrs}$ from the jet-driven shell simulation in \figref{fig:long_term:maps} 
    and \figref{fig:long_term:PV_transverse}, inclined by $i=\SI{55}{\degree}$ from the line of sight (connected black symbols).
    Blue and red squares show observed MV distributions in the blueshifted and redshifted lobes of the HH46-47 outflow, as derived by \cite{Zhang16}. Open squares assume optically thin $\ce{CO}$ emission, while filled squares include a velocity-dependent correction for optical depth (see text). The dashed lines show power-laws of slope $\gamma = -3.4$ (red) and $\gamma = -2.7$ (blue).
    }
    \label{fig:long_term:mv}
\end{figure}

{As an additional diagnostic tool, \figref{fig:long_term:mv} shows the mass-velocity (MV) distribution of entrained core material derived from the long term simulation in \figref{fig:long_term:maps} 
at an age $t=\SI{10000}{\yrs}$, assuming an inclination of $i=\SI{55}{\degree}$ from the line of sight.}
This MV distribution is confronted to that derived by \cite{Zhang16} from \ce{CO} observations of the HH46-47 outflow, which is viewed at the same inclination angle.

Two observed MV distributions were computed by \cite{Zhang16} and are plotted in \figref{fig:long_term:mv}: the first one (open symbols) was derived from $^{12}$CO without correction for optical depth. The shallow power-law slope $\gamma \simeq -2$ is very typical of the slopes reported previously in other CO outflows, before optical-depth correction. The second observed MV distribution (filled symbols) was obtained after applying a velocity-dependent optical-depth correction to $^{12}$CO and $^{13}$CO data, and adding (optically thin) C$^{18}$O data at low velocities. Because optical depth increases at lower flow velocity, the corrected distribution exhibits a much steeper slope $\gamma \simeq -3.4$ (in log-log). Similar steeper slopes have been obtained in other outflows after applying a velocity-dependent optical-depth correction, such as in B5-IRS1 \citep{Yu99}. 

{\Figureref{fig:long_term:mv} shows a remarkably good qualitative agreement between the simulated and the observed MV distribution in HH46-47 after correction for optical depth, which should be closest to the true mass distribution in the outflow. 
This agreement further reinforces the attractiveness of the jet-driven shell model at realistic long ages and with ambient stratification as a possible origin for CO outflows.}

\subsection{Model approximations}
\label{section:caveats}
Here, we briefly discuss the main physics left out in our simulations and how it might affect the results. First, we considered an equilibrium atomic cooling curve, without following the out-of-equilibrium ionization and cooling behind shock fronts. Therefore, we cannot compute realistic synthetic emission maps in atomic and ionic lines, whose flux is proportional to the electron density. This approximation is sufficient, however, for our purpose of determining the overall shape, dynamics, and mass distribution in the dense shells of cooled post-shock gas. We also neglected chemistry, noting that CO dissociation only affects the observed mass distribution above 20 \kmps \citep{Downes-Cabrit2003,Moraghan08} and our comparisons with observations are made at lower velocities. Such approximations allow us to carry out for the first time, in a cost-effective way, long-term simulations up to $10^4$ yr that can be compared with actual outflows observed with ALMA.

Overall, we neglected the infall motions. This allowed us to properly compare with simulations of outflows driven by X-winds into the same stratified singular cores \citep{LeeEtAl2001,Shang20}, where infall is likewise neglected. In such cores, infall propagates inside-out at the ambient sound speed, $a$, and generates a shallower density distribution, $1/r^{1.5}$, inside the sonic radius, $r_{\rm inf} = a t$ \citep[][]{TerebeyShuCassen84}.
Contrary to wind-driven flows where it can lead to a steady shell \citep{Liang20},
a $1/r^{1.5}$ slope does not greatly reduce the width of jet-driven flows, as compared to $1/r^{2}$ \citep{Moraghan08}. The main effect of infall would thus be to add shear, entrainment, and extra ram pressure at the base of the outer shell \citep[cf.][for the wind-driven case]{Liang20}. For our simulation parameters, infall would occur inside $r_{\rm inf} \simeq 1000~{\rm au}\, (a/0.5 \kmps)\times (t/10^4 yr)$ and, hence, it would affect only a small fraction of our full computational domain, extending up to 20,000 au.

Our simulation parameters do not explore the "long-period" regime where the ambient core would have time to partly refill the cavity in between jet outbursts. 
This interesting situation would occur if major outbursts happen only every few 10$^4$ yr, for instance, due to tidal interaction in wide eccentric binaries. The observed spacing of jet knots in Class 0/1 sources indicates much shorter variability timescales, however, with multiple periodic modes of decades, centuries, and thousands of years  \citep{Raga12,Lee2020}. In addition, a recent study of infrared variability towards embedded protostars with Spitzer/IRAC presented the conclusion that Class 0 protostars undergo a major burst on average every 438 yr, with a 95\% confidence interval of 161 to 1884 yr \citep{Zakri22}. The long-period regime thus seems quite rare among the youngest, embedded Class 0/1 objects that drive observable CO outflows.

Finally, we have neglected magnetic fields, both in the jet and in the ambient medium. Magnetic pressure in the jet would act to reduce the postshock compression and increase the cooling length, by typically an order of magnitude \citep{Hartigan94}. This effect is observed directly in resolved internal shocks of stellar jets \citep{Hartigan2015}. High-resolution radiative numerical simulations of pulsed magnetized jets show that the reduced cooling tends to broaden the nested bowshock shells, but the effect appears rather modest \citep{deColle06}, justifying its omission here.

Conversely, a magnetic field in the ambient medium would tend to resist against the sideways shell expansion. A significant reduction in shell width compared to the purely hydrodynamic case requires, however, strong fields near equipartition (i.e., a ratio of thermal to magnetic pressure $\beta \simeq 1$). This is demonstrated, for example, in the simulations of \citet{Shang20} of X-wind-driven shells into cores of varying degrees of flattening and magnetization\footnote{The ratio of thermal to magnetic pressure in the core models of \citealt{Shang20} can be recovered as $\beta = \alpha_b^{-2}\, (v_A/a)^{-2}$ where  $\alpha_b$ = [0, 0.1, 1] is their scaling parameter and $(v_A/a)$ is the ratio of Alfv\'enic speed to sound speed in the magnetostatic solution of \citealt{LiEtShu1996} (Eq. 69), which increases with the core flattening parameter $n$. A significant reduction in shell width compared to the hydrodynamical case is seen only in simulations with $n \ge 4$ and $\alpha_b = 1$, corresponding to midplane values of $\beta \le 3.5$ that approach equipartition}. For the standard moderate core flattening, $n = 2,$ adopted in the present work, the maximum reduction in shell width due to ambient magnetic field is only 25\%, therefore, our predicted cavity shapes and dynamics should remain valid. We note that an added complexity at higher core magnetization, in the general case of non-zero rotation, would be the probable launching of a massive slow MHD wind from the Keplerian disk formed around the protostar \citep[see e.g.,][and references therein]{Lesur21}. The interaction of an inner pulsed jet (or wide-angle X-wind) with an outer disk wind could significantly affect the formation of outflow cavities, but the long-term evolution has only been examined in the hydrodynamical case so far, to our knowledge \citep{Tabone18}.

Concerning observational predictions, we focused here on signatures of the shell shapes, kinematics, and mass distribution at velocities below 20 \kmps, which are well traced by low-excitation CO emission observable with ALMA. Our simulations may also be used to assess the detectability of warmer jet-originated material inside the cavity volume. From Figure~\ref{fig:long_term:maps} we estimate that the nested shells (driven by each of the jet pulses) generate typical shock speeds of 30~\kmps\ inside the cavity. The pre-shock density inside the cavity is $\simeq$ 100 cm$^{-3}$ at $z \le 3000$ au, and drops off at higher altitudes. Using the atomic shock model grid of \citet{Hartigan94}, the maximum predicted surface brightness in [S II] for solar abundances is then $\simeq 5 \times 10^{-6}$ erg s$^{-1}$ cm$^{-2}$ sr$^{-1}$. Such extended [S II] emission inside outflow cavities would be worthwhile to search for, but it might be difficult to isolate against the much brighter axial jet and scattered light from the cavity walls, especially since young protostars with powerful CO outflows are often located in regions of high optical extinction.

When the jet is dense enough to be partly molecular, another possible tracer of nested shells inside the main outflow cavity is H$_2$ ro-vibrational emission, the most spectacular example so far being the Class 0 outflow HH 212. Each large H$_2$ bowshock, produced by a major jet pulse, is seen to connect to a separate CO shell at the flow base, nested inside the main cavity \citep[see Figure 2 in][]{Lee2015}, a morphology consistent with our predictions for a pulsed jet-driven outflow. Modeling the H$_2$ line emission could be helpful to further discriminate between this scenario and the pulsed wide-angle wind model proposed by \citet{Zhang19}.

\section{Conclusions}

    We confirm in this paper that the swept-up shell driven by a cylindrical pulsed jet opens much wider inside an ambient core with steeply decreasing density, than within a homogeneous core (which was the configuration commonly adopted until now for analytical models and most simulations of jet-driven bowshocks). At early times of a few hundred years,  
    the jet-driven shell can open as wide as with a wide-angle X-wind, when considering the same flattened singular ambient core and the same injected mass-flux and velocity variability.
    
    Then, we investigated the impact that several parameters in our model have on the general morphology, opening angle, and kinematics of jet-driven shells. The parameter which is by far the most impactful is the ambient core density, followed by the jet density and jet opening angle in a less impactful fashion. Within the range of values explored in this paper, the other parameters do not significantly affect these diagnostics.
    
    Finally, running a representative simulation up to $t=\SI{10000}{\yrs}$ reveals drastic changes in the long-term. After several hundred years, the initial shell splits into a slow parabolic outer shell fully made of core-original material (tracing the forward shock) and an inner, faster main shell tracing the jet-ambient interface, which stops expanding at the base after $\SI{8000}{\yrs}$, unlike wide-angle wind-driven shells. This main shell encompasses a mixed-material cavity inside which successive bowshocks driven by the pulsed jet expand and slow down by interacting with previous ones, producing a strongly sheared velocity field parallel to the (roughly conical) main shell walls. Both the morphology and the velocity fields are very different from analytical predictions of ballistic jet bowshock models in uniform media \citep{Ostriker2001}. 
    
    The long-term simulation of our basic model shows none of the caveats of steady jet bowshocks in uniform media (excessive length to width ratio, excessive blue-red overlap, and overly low speeds at large shell widths). \mialyadd{On the contrary, it shows} very promising similarities with the most recent observations of Class 0 outflows observed at high resolution with ALMA, in terms of predicted shell widths, \mialyadd{full opening angle ($\simeq 90$\degr after $10^4$ yr),} longitudinal and transverse position-velocity cuts, and mass-velocity distribution \citep{Dutta2020,Plunkett15,Zhang16,Zhang19}. Some comparisons even show more resemblance to observations than the widely used wide angle "modified X-wind" model of \cite{LeeEtAl2001}. \mialyadd{This is the case for} the moderate outflow widths at $\SI{800}{au}$, the characteristic "bell-shape" of transverse PV cuts, and the steep mass-velocity relation after CO opacity correction.  

    More generally, this paper shows that a realistic modeling of the surrounding core density stratification, as well as long integration times of at least $10^4~\SI{}{\yrs}$, are both essential to reliably predict the properties of outflows driven by a pulsed jet, and to confront them with observations In the future, we plan to extend our simulations to include chemistry, magnetic field, rotation, and precession to model specific observed protostellar outflows in more detail.

\begin{acknowledgements}
We are grateful to the anonymous referee for insightful suggestions that improved the paper, and to Doug Johnstone for supportive discussions and thoughtful comments. We acknowledge support by the Programme National Physique et Chimie du Milieu Interstellaire (PCMI) of CNRS/INSU with INC/INP co-funded by CEA and CNES. Part of the computations  were carried out on the OCCIGEN cluster at CINES \footnote{\url{https://www.cines.fr/}} in Montpellier (project named \texttt{lut6216}, allocation \texttt{A0090406842} and \texttt{A0100412483}) and on the MesoPSL cluster\footnote{\url{http://www.mesopsl.fr/}} of PSL University at Observatoire de Paris. 
\end{acknowledgements}

% for the bibliography, at the end
\bibliographystyle{aa} % style aa.bst
\bibliography{main} % your references .bib

\begin{appendix}
\section{Effect of variables with lower jet density}
\label{appendix:lower_mass_flow_rate}

In this appendix, we redo the process of \Sectref{section:monovariated_run}, but starting from a reference model with a jet base density that is two lower by orders of magnitude:\ 
 $\rho_{j0}=\SI{1.8e-19}{\gram\per\centi\meter\cubed}$. In this way, we can further probe the effects of the base jet density on the morphology and kinematics of the driven shells. We also test the robustness of the conclusions we got from \Sectref{section:monovariated_run} at other jet density values.

Thus, additionally to the reference model, we launch seven modified models, each having the same parameters as the reference model except for a single one. Those models are
: 

\textit{(i)}~M\_DENSA: with an ambient density scaling $\rho_{a,0}$ of $\SI{1.6e-20}{\gram\per\centi\meter\cubed}$ instead of $\SI{1.6e-18}{\gram\per\centi\meter\cubed}$; 

\textit{(ii)}~M\_THETA: with a jet semi-opening angle $\theta_{j}$ of $\SI{7}{\degree}$ instead of $\SI{3}{\degree}$; 

\textit{(iii)}~M\_DENSJ: with a jet base density $\rho_{j0}$ 100 times larger than the reference case; 

\textit{(iv)}~M\_VARAMP: with a variability semi-amplitude $\Delta V$ of $\SI{90}{\kmps}$ instead of $\SI{60}{\kmps}$; 

\textit{(v)}~M\_PER: with a  variability period $P$ = $\SI{300}{\yrs}$ instead of $\SI{115}{\yrs}$; 

\textit{(vi)}~M\_SAWT: with a sawtooth instead of sinusoidal velocity variability profile ; 

\textit{(vii)}~M\_RAD: with an initial jet radius $\orthoradius_j$ of $\SI{20}{au}$ instead of $\SI{50}{au}$.

\begin{figure}[!th]
    \centering
    \resizebox{\hsize}{!}{\includegraphics[trim = 0 0.6cm 0 0 ,clip]{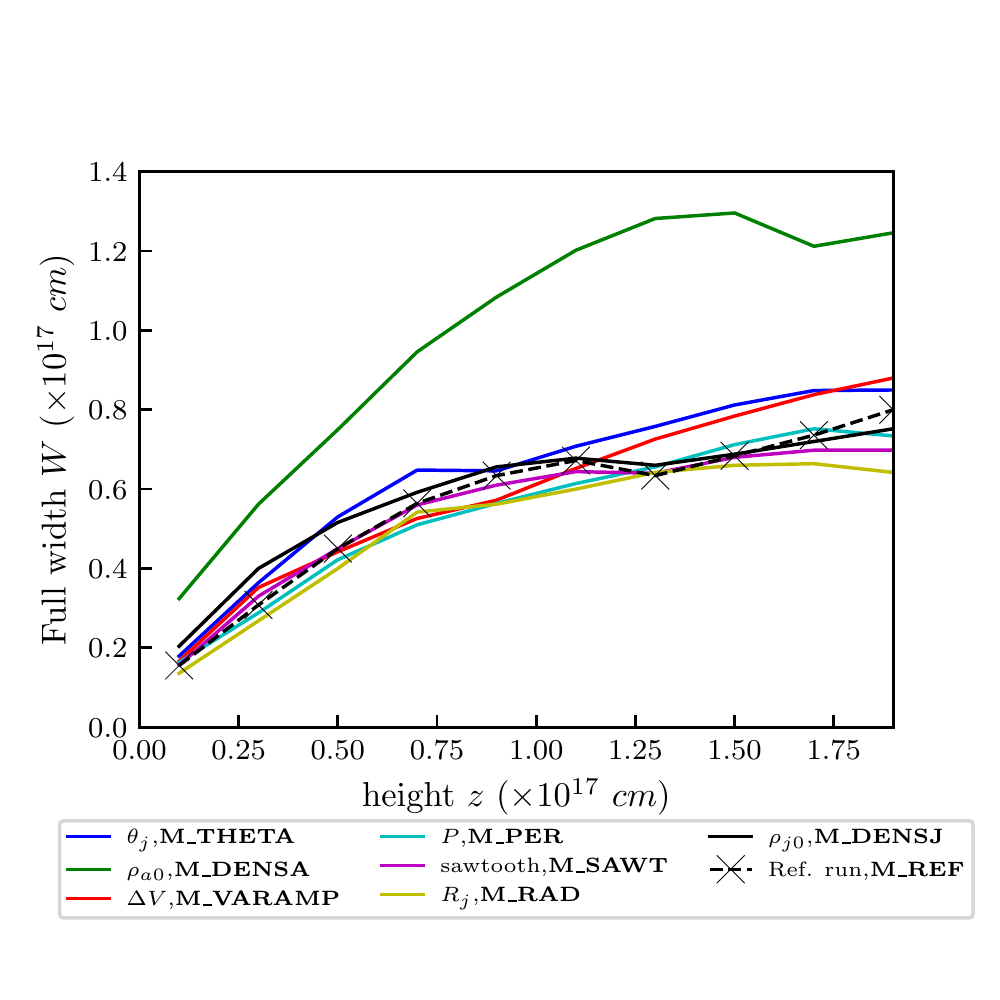}}
    \caption{Full shell width $\OpeningAngleWidth{}(z)$ at $t=\SI{700}{\yrs}$ as a function of altitude $z$ for the models in 
    \Tableref{table:monovariated_run:parameters_B}, with jet densities 100 times smaller than in \figref{fig:monovariated_run:full_width}. Colored curves have one parameter varied from the reference run, among the ambient core base density, $\rho_{a0}$, jet semi-opening angle ,$\theta_{j}$, initial jet base density, $\rho_{j0}$, semi-amplitude of variability, $\Delta V$, period, $P$, type of variability profile (sawtooth instead of sinusoidal), or jet radius, $\orthoradius_j$.}
    \label{fig:appendix:full_width}
\end{figure}

    \Figureref{fig:appendix:full_width} plots and compares the inferred shell shapes at $t=\SI{700}{\yrs}$ for the reference and modified models.
    \Tableref{table:monovariated_run:parameters_B} summarizes the parameters of the simulations, and lists the full opening angle at 800 au, $\alpha_{800}$, and full shell widths at $z=$ 800 au and 12700 au ($\OpeningAngleWidth{\SI{800}{}}$ and $\OpeningAngleWidth{\SI{12700}{}}$) measured at $t=\SI{700}{\yrs}$ in each model. The effect of parameters changes on the outflow structure are discussed in \Sectref{section:monovariated_run:global_results}.

\begin{table*}
\caption{Parameters of conical pulsed medium-density jet simulations with the resulting opening angles and full widths
}
\label{table:monovariated_run:parameters_B}
\begin{center}
\begin{minipage}{\textwidth}
\begin{tabular}{l l l l l l l}
 \toprule
  \hline
     \multicolumn{7}{c}{\textit{Fixed parameters}}
     \\
     \hline
     mean jet velocity  &
     \multicolumn{6}{l}{$v_0 = \SI{120}{\kmps}$}
     \\
     jet density variation  &
     \multicolumn{6}{l}{$\rho_j(t) = \rho_{j0}\,[v_0/ v_j(t)]\times(\orthoradius_j^2+z_0^2)(\orthoradius^2+[z+z_0]^2)^{-1}$, with $z_0=\orthoradius_j/\tan\theta_j$\quad (constant mass-flux)}
     \\ 
     jet temperature  &
     \multicolumn{6}{l}{$T_j$ = $\SI{100}{\kelvin}$}
     \\
     ambient core temperature  &
     \multicolumn{6}{l}{$T_a = \SI{100}{\kelvin}$}
     \\
     core density profile & \multicolumn{6}{l}{Flattened singular core
     $\rho_a(\mathbf{r})=\rho_{a0}\sin^2{\theta} \,{\radius_0^2}\,\left({\radius^2}\right)^{-1}$, with $\radius_0=\SI{2.5e15}{\cm}$}
     \\
     radiative cooling function  &
     \multicolumn{6}{l}{$\Lambda\left(\SI{100}{\Kelvin}\leq T <10^4~\SI{}{\Kelvin}\right)$ from \cite{DalgarnoEtMcCray72}}
     \\
     & 
     \multicolumn{6}{l}{$\Lambda\left(T\geq10^4~\SI{}{\Kelvin}\right)$ from \cite{Schure09}}
     \\
     simulation domain  &
     \multicolumn{6}{l}{$\left(\orthoradius,z\right)
     =
     \left(\SI{7.0e16}{\cm},\SI{1.9e17}{\cm}\right)=\left(\SI{4679}{au},\SI{12700}{au}\right)$}
     \\
     number of cells  &
     \multicolumn{6}{l}{$n_\orthoradius\times n_z
     =
     56\times152$ at AMR level 1}
     \\
     maximal AMR level & 5 \\
          maximum resolution  &
     \multicolumn{6}{l}{$\Delta\orthoradius=\Delta z=\SI{7.8e+13}{\cm}=\SI{5.2}{au}$ at AMR level 5}
     \\
     snapshot age  & \multicolumn{6}{l}{$\SI{700}{\yrs}$} 
     \\
     \toprule\hline
       \multirow{2}{1.0cm}{Parameter} & 
       \multirow{2}{2.5cm}{Reference Model M\_REF} & 
       \multirow{2}{2.1cm}{Modified parameter 
       \footnote{In each modified model, only one parameter at a time is changed with respect to the reference model.}}&
       \multirow{2}{2.2cm}{Modified Model name}
      & $\alpha_{800}$ \footnote{Shell full opening angle at $z = 800$ au. We obtain $\alpha_{800}=\SI{72}{\degree}$ for the reference model.}
      & $\OpeningAngleWidth{\SI{800}{}}$\footnote{Full shell width at $z = 800$ au. We obtain $\OpeningAngleWidth{\SI{800}{}}=\SI{1.7e16}{\cm}$ for the reference model.} 
      & 
      $\OpeningAngleWidth{\SI{12700}{}}$\footnote{Full shell width at $z = 12700$ au (top of the grid). We obtain $\OpeningAngleWidth{\SI{12700}{}}=\SI{8.0e16}{\cm}$ for the reference model.}
     \\ 
        & 
        & 
       &  
      & ($\SI{}{\degree}$)
      & ($10^{16}~\SI{}{\cm}$)
      & ($10^{16}~\SI{}{\cm}$)
     \\ 
     \hline
     core base density $\rho_{a0}$ &
     $\SI{1.6e-18}{\gram\per\centi\meter\cubed}$ &      $\SI{1.6e-20}{\gram\per\centi\meter\cubed}$ 
     &
     M\_DENSA &  $\SI{112}{}$
     & $\SI{3.5}{}$
     & $\SI{12.4}{}$
     \\
     jet semi opening angle  $\theta_j$ &
     $\SI{3}{\degree}$ & $\SI{7}{\degree}$ &
     M\_THETA & $\SI{81}{}$
     & $\SI{2.0}{}$
     & $\SI{8.4}{}$ \\
       jet base initial density\footnote{Medium jet density case, yielding a one-sided mass-flux 
       $\massflowrate = \SI{6e-8}{\MSun/yr}$ for the reference values of $\theta_j$ and $R_j$}  $\rho_{j0}$&
      $\SI{1.8e-19}{\gram\per\centi\meter\cubed}$ & $\SI{1.8e-17}{\gram\per\centi\meter\cubed}$ &
     M\_DENSJ\footnote{This model is equivalent to H\_REF.} & $\SI{88}{}$ 
     & $\SI{2.3}{}$
     & $\SI{7.6}{}$
     \\    
     semi-amplitude $\Delta V$ &
     $\SI{60}{\kmps}$ & $\SI{90}{\kmps}$ &
     M\_VAMP & $\SI{77}{}$ 
     & $\SI{1.9}{}$
     & $\SI{8.8}{}$
     \\
     jet variability period $P$  &
     $\SI{115}{\yrs}$ & 
     $\SI{300}{\yrs}$ &
     M\_PER &  $\SI{74}{}$ 
     & $\SI{1.8}{}$
     & $\SI{7.4}{}$\\ 
     jet variability profile $h(t)$ &
      $1+ \frac{\Delta V}{v_0} \sin {\frac{2\pi t}{P}}$ & 
      $1+\frac{\Delta V}{v_0}
      \left\{1 - \right.$
      & \multirow{2}{0.5cm}{M\_SAWT} & \multirow{2}{0.5cm}{$\SI{74}{}$} 
      & \multirow{2}{0.5cm}{$\SI{1.8}{}$}
      & \multirow{2}{0.5cm}{$\SI{7.0}{}$}
     \\
     &
       & 
      $\left.2\cdot \mathrm{mod}\left(\frac{t}{P},1\right)\right\}$
      &  & &
     \\
     jet radius $\orthoradius_j$ &
      $\SI{7.5e14}{\cm}$ & $\SI{3.0e14}{\cm}$
       & 
     M\_RAD & $\SI{65}{}$ 
     & $\SI{1.5}{}$
     & $\SI{6.4}{}$
     \\
\toprule
\end{tabular}
\end{minipage}
\end{center}
\end{table*}

\end{appendix}

\end{document}